\documentclass[reprint,showpacs,preprintnumbers,nofootinbib,amsmath,amssymb,aps,prd,floatfix,superscriptaddress]{revtex4-1}
\pdfoutput=1
\usepackage{graphicx}
\usepackage{bm}
\usepackage{subfigure}
\usepackage{url}
\usepackage[hyperindex]{hyperref}
\usepackage{color}
\usepackage[ddmmyy,24hr]{datetime}
\usepackage{bigdelim}
\usepackage{booktabs}
\usepackage{dcolumn}
\usepackage{multirow}
\usepackage{subfigure}
\usepackage{cancel}
\usepackage{stackrel}
\usepackage{paralist}
\usepackage{xspace}
\usepackage{slashed}
\newcommand{\nua}[1]{\ensuremath{\rlap{\kern-2.5pt\ensuremath{\overset{\scriptscriptstyle(-)}{\phantom{\nu}}}}{\ensuremath{{\nu}_{#1}}}}}
\newcommand{\vet}[1]{\ensuremath{\hskip-1pt\vec{\hskip1pt#1}}}
\usepackage{enumitem}

\begin{document}

\title{Neutrino Charge Radii from COHERENT Elastic Neutrino-Nucleus Scattering}

\author{M. Cadeddu}
\email{matteo.cadeddu@ca.infn.it}
\affiliation{Dipartimento di Fisica, Universit\`{a} degli Studi di Cagliari,
and
INFN, Sezione di Cagliari,
Complesso Universitario di Monserrato - S.P. per Sestu Km 0.700,
09042 Monserrato (Cagliari), Italy}

\author{C. Giunti}
\email{carlo.giunti@to.infn.it}
\affiliation{Istituto Nazionale di Fisica Nucleare (INFN), Sezione di Torino, Via P. Giuria 1, I--10125 Torino, Italy}

\author{K.A. Kouzakov}
\email{kouzakov@gmail.com}
\affiliation{Department of Nuclear Physics and Quantum
Theory of Collisions, Faculty of Physics, Lomonosov Moscow State University, Moscow 119991, Russia}

\author{Y.F. Li}
\email{liyufeng@ihep.ac.cn}
\affiliation{Institute of High Energy Physics,
Chinese Academy of Sciences, Beijing 100049, China}
\affiliation{School of Physical Sciences, University of Chinese Academy of Sciences, Beijing 100049, China}

\author{A.I. Studenikin}
\email{studenik@srd.sinp.msu.ru}
\affiliation{Department of Theoretical Physics, Faculty of
Physics, Lomonosov Moscow State University, Moscow 119991, Russia}
\affiliation{Joint Institute for Nuclear Research, Dubna 141980, Moscow Region, Russia}%

\author{Y.Y. Zhang}
\email{zhangyiyu@ihep.ac.cn}
\affiliation{Institute of High Energy Physics,
Chinese Academy of Sciences, Beijing 100049, China}
\affiliation{School of Physical Sciences, University of Chinese Academy of Sciences, Beijing 100049, China}

\date{26 February 2020}

\begin{abstract}
\hfill\textbf{Revised version of Phys. Rev. D 98 (2018) 113010}\hfill\null
\\
Coherent elastic neutrino-nucleus scattering
is a powerful probe of neutrino properties,
in particular of the neutrino charge radii.
We present the bounds on the neutrino charge radii obtained from the analysis of the
data of the COHERENT experiment.
We show that the time information of the COHERENT data
allows us to restrict the allowed ranges of the neutrino charge radii.
We also obtained for the first time bounds on the
neutrino transition charge radii,
which are quantities beyond the Standard Model.
\end{abstract}


\maketitle

\section{Introduction}
\label{sec:introduction}

Neutrinos are widely believed to be neutral particles,
but in reality they could have a very small electrical charge
and it is very likely that they have charge radii
(see the review in Ref.~\cite{Giunti:2014ixa}).
Indeed,
in the Standard Model neutrinos have charge radii of the order of
$10^{-33} \, \text{cm}^{2}$
\cite{Lee:1973fw,Lee:1977tib,Lucio:1983mg,Lucio:1984jn,Degrassi:1989ip,Papavassiliou:1989zd,Bernabeu:2000hf,Bernabeu:2002nw,Bernabeu:2002pd,Fujikawa:2003ww,Papavassiliou:2003rx,Bernabeu:2004jr}.
In this paper we consider the effects of the neutrino charge radii
on coherent elastic neutrino-nucleus scattering
\cite{Sehgal:1985iu,Papavassiliou:2005cs,Kosmas:2015vsa,Kosmas:2017tsq}
and we present the results on the values of the neutrino charge radii obtained
from the analysis of the data of the COHERENT experiment~\cite{Akimov:2017ade,Akimov:2018vzs}.

Coherent elastic neutrino-nucleus scattering is a process
predicted a long time ago~\cite{Freedman:1973yd,Freedman:1977xn,Drukier:1983gj},
which was observed for the first time
in 2017 in the COHERENT experiment~\cite{Akimov:2017ade,Akimov:2018vzs}.
The difficulty is that it is necessary to observe
nuclear recoils with very small kinetic energy $T$,
smaller than a few keV,
in order to satisfy the coherence requirement
$|\vec{q}| R \ll 1$~\cite{Bednyakov:2018mjd},
where $|\vec{q}| \simeq \sqrt{2 M T}$ is the three-momentum transfer,
$R$ is the nuclear radius of a few fm,
and
$M$ is the nuclear mass,
of the order of 100 GeV for heavy nuclei.
The observation of coherent elastic neutrino-nucleus scattering
opens up a new and powerful way to probe the properties of
nuclei, neutrinos, weak interactions, and new physics beyond the Standard Model
\cite{Barranco:2005yy,Patton:2012jr,Papoulias:2015vxa,Lindner:2016wff,Shoemaker:2017lzs,Ciuffoli:2018qem,Canas:2018rng,Billard:2018jnl,Brdar:2018qqj}.
Indeed,
the first measurements of the COHERENT experiment have already produced interesting results for
nuclear physics~\cite{Cadeddu:2017etk},
neutrino properties and interactions~\cite{Coloma:2017ncl,Liao:2017uzy,Kosmas:2017tsq,Denton:2018xmq,AristizabalSierra:2018eqm}, and
weak interactions~\cite{Cadeddu:2018izq}.

The problem of correctly defining the neutrino charge radius in the context
of the Standard Model and beyond has a long history
(see the review in Ref.~\cite{Giunti:2014ixa}).
The authors of one of the first studies \cite{Bardeen:1972vi}
found that in the Standard Model and in the unitary gauge the neutrino charge radius
is ultraviolet divergent and, hence, not a physical quantity.
A direct one-loop calculation
\cite{Dvornikov:2003js,Dvornikov:2004sj} of the neutrino charge radius,
accounting for contributions
of a complete set of proper vertexes and $\gamma-Z$
self-energy Feynman diagrams,  performed in a general $R_{\xi}$ gauge for a massive Dirac neutrino,
also gave a divergent result.
The solution to the problem of obtaining a charge radius
that is gauge-independent, finite and independent of the external probe
(see Ref.~\cite{Papavassiliou:1989zd} for a detailed discussion)
is achieved by including appropriate additional diagrams
in the calculation of the neutrino electromagnetic form factor.
In the usual approach, the Feynman diagrams are treated individually,
and each diagram either contributes to the form factor in its entirety
or it does not contribute at all.
However, this method yields an infinite and gauge-dependent charge radius.
The problem is that certain diagrams,
which at first glance do not appear to be relevant for the calculation of the form factor,
contain pieces that cannot be distinguished from the contributions of the regular diagrams and
must therefore be included in order to obtain a finite result.
The appropriate way to include those diagrams,
found in Ref.~\cite{Papavassiliou:1989zd},
is based on the pinch technique.
The resulting neutrino charge radii are
finite and independent of the gauge and the external
probe~\cite{Bernabeu:2000hf,Bernabeu:2002nw,Bernabeu:2002pd}.

Until now, the neutrino charge radii have been typically searched in elastic
neutrino-electron scattering experiments.
Summaries of the limits
obtained so far in this way can be found in Refs.~\cite{Giunti:2014ixa,Tanabashi:2018oca}
and in Table~\ref{tab:limits} of this paper.
For small energy transfer $T$,
both the Standard Model cross section
and the effect of the neutrino charge radii in the case of elastic
neutrino-electron scattering turn out to be
smaller by a factor of the order of $M/m_e$
with respect to the case of coherent elastic neutrino-nucleus scattering.
Therefore, in terms of data collection,
coherent elastic neutrino-nucleus scattering experiments
have a greater potential for investigating the neutrino
charge radii than the measurements of neutrino-electron scattering.

In this paper we calculate accurately the limits on the neutrino charge radii
from the analysis of the COHERENT data,
starting with a discussion of the theoretical framework in Section~\ref{sec:framework},
which includes also a summary of the previous experimental limits
in Tab.~\ref{tab:limits}
and a discussion of other limits obtained with combined analyses of the data of different experiments.
In Sections~\ref{sec:spectrum} and \ref{sec:time}
we present, respectively, the results obtained from the analyses of the
time-integrated COHERENT energy spectrum
and
the time-dependent COHERENT data.
In Section~\ref{sec:conclusions} we draw our conclusions.

\begin{table*}
\footnotetext{\label{f2}Corrected by a factor of two due to a different convention.}
\footnotetext{\label{E734}Corrected in Ref.~\cite{Hirsch:2002uv}.}
\renewcommand{\arraystretch}{1.2}
\begin{tabular}{llcll}
Process & Collaboration & Limit [$10^{-32} \, \text{cm}^2$] & CL & Ref.\\
\hline
\multirow{2}{*}{Reactor $\bar\nu_{e}$-$e$}
&Krasnoyarsk	&$|\langle{r_{\nu_{e}}^{2}}\rangle|<7.3$	&90\%	&\cite{Vidyakin:1992nf}\\
&TEXONO		&$-4.2<\langle{r_{\nu_{e}}^{2}}\rangle<6.6$	&90\%	&\cite{Deniz:2009mu}\textsuperscript{\ref{f2}}\\
\hline
\multirow{2}{*}{Accelerator $\nu_{e}$-$e$}
&LAMPF		&$-7.12<\langle{r_{\nu_{e}}^{2}}\rangle<10.88$	&90\%	&\cite{Allen:1992qe}\textsuperscript{\ref{f2}}\\
&LSND		&$-5.94<\langle{r_{\nu_{e}}^{2}}\rangle<8.28$	&90\%	&\cite{Auerbach:2001wg}\textsuperscript{\ref{f2}}\\
\hline
\multirow{2}{*}{Accelerator $\nu_{\mu}$-$e$ and $\bar\nu_{\mu}$-$e$}
&BNL-E734	&$-5.7<\langle{r_{\nu_{\mu}}^{2}}\rangle<1.1$	&90\%	&\cite{Ahrens:1990fp}\textsuperscript{\ref{f2},\ref{E734}}\\
&CHARM-II	&$|\langle{r_{\nu_{\mu}}^{2}}\rangle|<1.2$	&90\%	&\cite{Vilain:1994hm}\textsuperscript{\ref{f2}}\\
\hline
\end{tabular}
\caption{\label{tab:limits}
Experimental limits for the neutrino charge radii.
}
\end{table*}

\section{Theoretical framework}
\label{sec:framework}

In the fundamental theory of electromagnetic neutrino interactions
the neutrino charge radii are defined for the massive neutrinos
(see the review in Ref.~\cite{Giunti:2014ixa}).
However,
the effects of neutrino oscillations can be neglected for experiments
with a short distance between the neutrino source and detector,
as in the setup of the COHERENT experiment.
In this case one can consider the effective charge radius
$\langle{r}_{\nu_{\ell}}^2\rangle$
of a flavor neutrino $\nu_{\ell}$,
with $\ell=e,\mu,\tau$.
Since in the ultrarelativistic limit the charge form factor conserves the neutrino helicity,
as the Standard Model weak interactions,
the contribution of $\langle{r}_{\nu_{\ell}}^2\rangle$
to the elastic scattering of $\nu_{\ell}$
with a charged particle adds coherently to the Standard Model weak interactions
and can be expressed through the shift
\cite{Degrassi:1989ip,Vogel:1989iv,Kouzakov:2017hbc}
\begin{equation}
\sin^2\!\vartheta_{W}
\to
\sin^2\!\vartheta_{W}
\left(
1
+
\frac{1}{3} m_{W}^{2}
\langle{r}_{\nu_{\ell}}^2\rangle
\right)
,
\label{shift}
\end{equation}
where $\vartheta_{W}$ is the weak mixing angle,
$m_{W}$ is the mass of the $W$ boson,
and $\ell=e,\mu,\tau$.
This shift follows from the expression\footnote{
For simplicity we omitted a term
$ q_{\mu} \slashed{q} / q^{2} $
whose contribution vanishes
in the coupling with a conserved current as in
neutrino-electron and neutrino-nucleon scatterings.}
\begin{equation}
\Lambda_{\mu}^{(\nu)}(q)
=
\gamma_{\mu}
F_{\nu}(q^{2})
\simeq
\gamma_{\mu} q^{2} \, \frac{\langle{r}^2\rangle}{6}
\label{vertex}
\end{equation}
of the effective electromagnetic interaction vertex
of neutrinos in the Standard Model,
where $F_{\nu}(q^{2})$ is a form factor,
$q$ is the four-momentum transfer
and the approximation holds for small values of $q^2$.
The charge radius is given by
(see Ref.~\cite{Giunti:2014ixa})
\begin{equation}
\langle{r}^2\rangle = 6 \left. \frac{d F_{\nu}(q^{2})}{d q^2} \right|_{q^2=0}
.
\label{chrdef}
\end{equation}

Unfortunately
in the literature there is some confusion on the size and the sign of the shift
of $\sin^2\!\vartheta_{W}$ due to a neutrino charge radius.
The authors of Ref.~\cite{Hirsch:2002uv}
considered a shift that has the same magnitude but opposite sign.
The authors of Refs.~\cite{Barranco:2007ea,Kosmas:2017tsq}
considered a shift that is twice as large,
with the same sign,
which corresponds to values of the neutrino charge radii which are half of ours.
The authors of Ref.~\cite{Grau:1985cn,Papavassiliou:2005cs}
considered a shift that is twice as large,
with opposite sign.
This implies that the Standard Model values of the neutrino charge radii
reported in Ref.~\cite{Papavassiliou:2005cs},
according to the calculations in Refs.~\cite{Bernabeu:2000hf,Bernabeu:2002nw,Bernabeu:2002pd},
are half and with opposite sign with respect to those that would be obtained in our framework.
We think that the sign of the shift can be considered as a convention
on the definition of $\langle{r}^2\rangle$
as
$ \pm 6 \, d F_{\nu}(q^{2}) / d q^2 |_{q^2=0} $.
Indeed,
in Ref.~\cite{Bernabeu:2000hf} it is explicitly written that
$\langle{r}^2\rangle = - 6 \, d F_{\nu}(q^{2}) / d q^2 |_{q^2=0} $,
which differs by a sign from our definition in Eq.~(\ref{chrdef}).
We think that the difference of a factor of two is due to the
assumption of a contribution to the effective electromagnetic interaction vertex
of an anapole moment with the same value of the charge radius,
which leads to a doubling of the shift of $\sin^2\!\vartheta_{W}$.
This is indicated by Eq.~(8)\footnote{Eq.~(2.8) in the arXiv version.}
of Ref.~\cite{Bernabeu:2002pd},
where the $\gamma^{5}$ term is due to an anapole moment assumed to have the same value
as the charge radius~\cite{Bernabeu:2002nw}.
Acting on left-handed spinors with $ 1 - \gamma^{5} = 2 $
leads to the doubling of the shift
of $\sin^2\!\vartheta_{W}$.
However,
as explained in Ref.~\cite{Giunti:2014ixa}
this approach is not well justified because in the Standard Model there is only
the form factor in Eq.~(\ref{vertex}),
which can be interpreted either as a charge radius or as an anapole moment\footnote{
Since the anapole moments have the same effects on the interactions of ultrarelativistic neutrinos
as the corresponding charge radii,
the phenomenological constraints on the charge radii
apply also to the anapole moments
(multiplied by $-6$ in the conventions of Ref.~\cite{Giunti:2014ixa}).
}.
Taking into account these considerations,
in our framework the Standard Model predictions of the neutrino charge radii
calculated in Refs.~\cite{Bernabeu:2000hf,Bernabeu:2002nw,Bernabeu:2002pd}
are given by
\begin{equation}
\langle{r}_{\nu_{\ell}}^2\rangle_{\text{SM}}
=
-
\frac{G_{\text{F}}}{2\sqrt{2}\pi^{2}}
\left[
3-2\ln\left(\frac{m_{\ell}^{2}}{m^{2}_{W}}\right)
\right]
,
\label{G050}
\end{equation}
where $m_{W}$ and $m_{\ell}$ are the $W$ boson and charged lepton masses
($\ell=e,\mu,\tau$).
The shift of $\sin^2\!\vartheta_{W}$
given by this expression of $\langle{r}_{\nu_{\ell}}^2\rangle_{\text{SM}}$
and Eq.~(\ref{shift})
is in agreement with the main contribution calculated in
Refs.~\cite{Sakakibara:1979rc,Sehgal:1985iu}
and with the shift given in Ref.~\cite{Erler:2013xha}.
Numerically, we have
\begin{align}
\null & \null
\langle{r}_{\nu_{e}}^2\rangle_{\text{SM}}
=
- 0.83 \times 10^{-32} \, \text{cm}^{2}
,
\label{reSM}
\\
\null & \null
\langle{r}_{\nu_{\mu}}^2\rangle_{\text{SM}}
=
- 0.48 \times 10^{-32} \, \text{cm}^{2}
,
\label{rmSM}
\\
\null & \null
\langle{r}_{\nu_{\tau}}^2\rangle_{\text{SM}}
=
- 0.30 \times 10^{-32} \, \text{cm}^{2}
.
\label{rtSM}
\end{align}
The current experimental bounds on
$\langle{r}_{\nu_{e}}^2\rangle$
and
$\langle{r}_{\nu_{\mu}}^2\rangle$
are listed in Tab.~\ref{tab:limits},
which is a corrected version of Tab.~V of Ref.~\cite{Giunti:2014ixa}.

The global fit of low-energy $\nu_{e}$-$e$ and $\bar\nu_{e}$-$e$ measurements
presented in Ref.~\cite{Barranco:2007ea} yielded the 90\% CL allowed interval
\begin{equation}
-0.26 \times 10^{-32}
<
\langle{r_{\nu_e}^{2}}\rangle
<
6.64 \times 10^{-32} \, \text{cm}^2
,
\label{BMR}
\end{equation}
where we have rescaled Eq.~(8) of Ref.~\cite{Barranco:2007ea}
taking into account the factor of two difference in the definition of the charge radii.
This range excludes
the Standard Model value of $\langle{r}_{\nu_{e}}^2\rangle$ in Eq.~(\ref{reSM}).
However,
we think that the allowed range in Eq.~(\ref{BMR})
must be corrected,
because it has been obtained assuming the
on-shell Standard Model value $\sin^2\!\vartheta_{W}^{\text{on-shell}} = 0.2227 \pm 0.0004$
obtained from a fit of high-energy electroweak measurements that do not involve neutrino-nucleon scattering
\cite{Zeller:2001hh}
and
considering the $\nu_{e}$--$e$ coupling
$
g_{V}^{\nu_{e}e}
=
\frac{1}{2} + 2 \sin^2\!\vartheta_{W}^{\text{on-shell}}
=
0.9454 \pm 0.0008
$.
At low energies the effective $\nu_{e}$--$e$ coupling is given by~\cite{Bahcall:1995mm}
\begin{equation}
g_{V}^{\nu_{e}e}
=
1 + \rho_{\nu e} \left( - \frac{1}{2} + 2 \hat{\kappa}_{\nu e} \hat{s}^2_{Z} \right)
,
\label{g_v_nue_e}
\end{equation}
where
$\rho_{\nu e} = 1.0126 \pm 0.0016$,
$\hat{\kappa}_{\nu e} = 0.9791 \pm 0.0025$,
and 
$\hat{s}^2_{Z}$ is the value of $\sin^2\!\vartheta_{W}$ at the $Z$ pole in the $\overline{\text{MS}}$ renormalization scheme.
From the LEP measurements at the $Z$ pole, which do not involve neutrino-nucleon scattering,
$\sin^2\!\vartheta_{W}^{\text{on-shell}} = 0.22331 \pm 0.00062$
\cite{ALEPH:2005ab}.
The corresponding value of $\hat{s}^2_{Z}$ is given by
$\hat{s}^2_{Z} = (1.0348 \pm 0.0002) \sin^2\!\vartheta_{W}^{\text{on-shell}}$
\cite{Tanabashi:2018oca},
which leads to
$\hat{s}^2_{Z} = 0.2311 \pm 0.0006$
and,
using Eq.~(\ref{g_v_nue_e}),
$g_{V}^{\nu_{e}e} = 0.952 \pm 0.002$.
Hence, the limits in Eq.~(8) of Ref.~\cite{Barranco:2007ea}
must be shifted by $ - ( 0.14 \pm 0.04 ) \times 10^{-32} \, \text{cm}^2$.
Adding in quadrature the uncertainties,
we obtained the 90\% CL allowed interval
\begin{equation}
-0.54 \times 10^{-32}
<
\langle{r_{\nu_e}^{2}}\rangle
<
6.37 \times 10^{-32} \, \text{cm}^2
.
\label{BMRcorrected}
\end{equation}
This allowed interval still excludes
the Standard Model value of $\langle{r}_{\nu_{e}}^2\rangle$ in Eq.~(\ref{reSM}),
but less strongly than the interval in Eq.~(\ref{BMR}).
We think that this tension requires further investigations,
that will be carried out elsewhere.

Constraints on $\langle{r_{\nu_{\mu}}^{2}}\rangle$
have been obtained in Ref.~\cite{Hirsch:2002uv}
from a reanalysis of the CCFR~\cite{McFarland:1997wx} and CHARM-II~\cite{Vilain:1994hm} data
on $\nu_{\mu}$-$e$ and $\bar\nu_{\mu}$-$e$ scattering.
Taking into account the sign difference in the definition of the charge radii,
in our framework the 90\% allowed interval in Eq.~(4.7) of Ref.~\cite{Hirsch:2002uv} becomes
\begin{equation}
-0.68 \times 10^{-32}
<
\langle{r_{\nu_{\mu}}^{2}}\rangle
<
0.52 \times 10^{-32} \, \text{cm}^2
.
\label{HNR}
\end{equation}
The Standard Model value of $\langle{r_{\nu_{\mu}}^{2}}\rangle$ in Eq.~(\ref{rmSM})
is within this interval.
The closeness of its value to the lower limit indicates that
future experiments may be able to measure $\langle{r_{\nu_{\mu}}^{2}}\rangle$.

The prescription in Eq.~(\ref{shift})
takes into account the contributions to neutrino interactions of
the charge radii of the three flavor neutrinos
$\nu_{e}$,
$\nu_{\mu}$,
$\nu_{\tau}$.
These are the only charge radii that exist in the Standard Model,
because the generation lepton numbers are conserved.
However,
in theories beyond the Standard Model
neutrinos can have transition charge radii
$\langle{r}_{\nu_{\ell\ell'}}^2\rangle$
that change the neutrino flavor.
For example,
in massive neutrino theories
the charge radii are defined in the mass basis of the physically propagating neutrinos.
The charge radii
$\langle{r}_{\nu_{\ell\ell'}}^2\rangle$
in the flavor basis are related to the charge radii
$\langle{r}_{\nu_{jk}}^2\rangle$
in the mass basis by the relation~\cite{Kouzakov:2017hbc}
\begin{equation}
\langle{r}_{\nu_{\ell\ell'}}^2\rangle
=
\sum_{j,k} U_{\ell j}^{*} U_{\ell' k} \langle{r}_{\nu_{jk}}^2\rangle
,
\label{basis}
\end{equation}
where $U$ is the neutrino mixing matrix.
Therefore,
even if the matrix of the neutrino charge radii is diagonal in the mass basis,
transition charge radii are generated by the mixing.

\begin{figure}[t!]
\begin{center}
\includegraphics*[width=\linewidth]{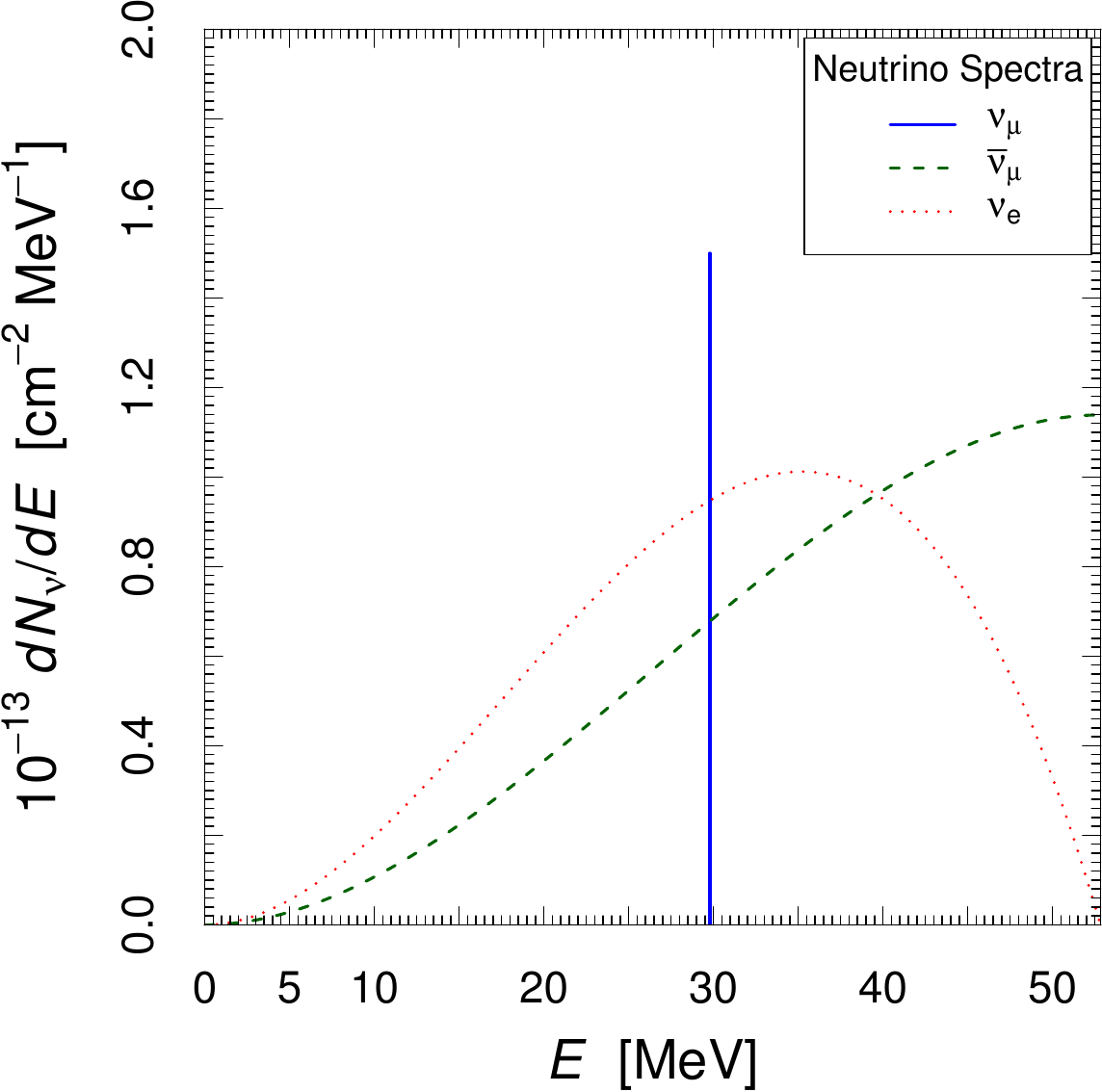}
\end{center}
\caption{ \label{fig:spec}
The COHERENT
$\nu_{\mu}$,
$\bar\nu_{\mu}$, and
$\nu_{e}$
spectra.
}
\end{figure}

\begin{figure*}[t!]
\begin{center}
\includegraphics*[width=\linewidth]{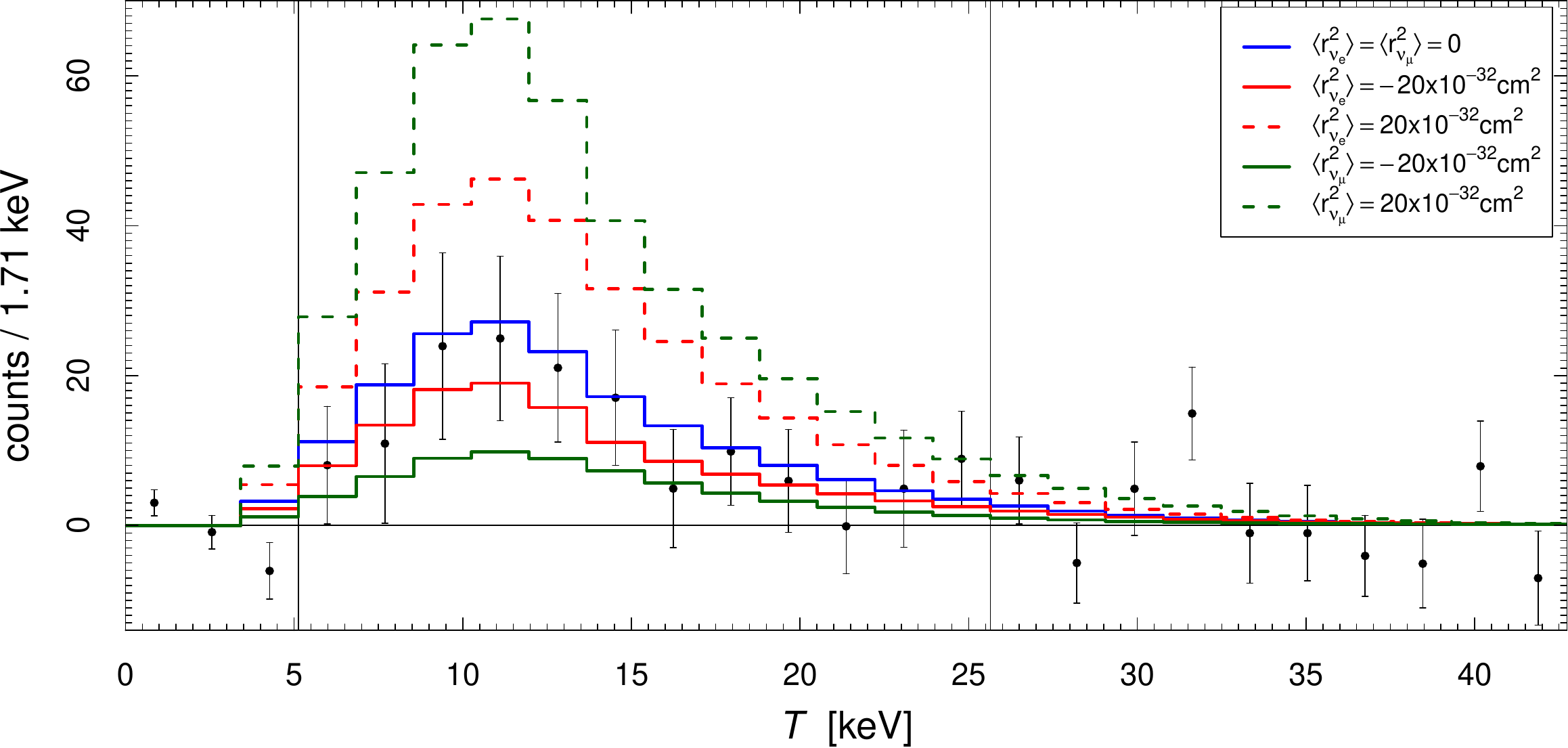}
\end{center}
\caption{ \label{fig:histT0}
COHERENT data \cite{Akimov:2017ade,Akimov:2018vzs} versus the nuclear kinetic recoil energy $T$.
The blue histogram shows the theoretical predictions
without neutrino charge radii and the theoretical values
of the rms radii of the neutron distributions in Eqs.~(\ref{RnCs}) and (\ref{RnI}).
The red and green histograms show the theoretical predictions obtained
by adding the contributions of the neutrino charge radii reported in the legend.
The two vertical lines at
5.13 and 25.64 keV indicate the range of our analysis, as explained after Eq.~(\ref{chi}).
}
\end{figure*}

The effects of the transition charge radii
$\langle{r}_{\nu_{\ell\ell'}}^2\rangle$,
was discussed for the first time in Ref.~\cite{Kouzakov:2017hbc}
considering the case of elastic neutrino-electron scattering.
Since the transition charge radii change the flavor of the neutrino in the final state
of the elastic scattering process,
the final state does not interfere with the weak interaction
channel and
the transition charge radii contributions add to the cross section
incoherently with respect to the standard weak interaction contribution.
In the case of coherent\footnote{
One should not confuse the meaning of the word ``coherent''
in ``coherent elastic neutrino-nucleus scattering''
with the coherency or incoherency of the charge radii contributions
with respect to the standard weak interactions.
The first coherency refers to the response of the nucleus as a whole to the
interaction,
whereas the second coherency refers to the interference of the final neutrino state.
}
elastic neutrino-nucleus scattering we consider the process
\begin{equation}
\nu_{\ell} + \mathcal{N} \to \sum_{\ell'} \nu_{\ell'} + \mathcal{N}
,
\label{cenns2}
\end{equation}
where $\mathcal{N}$ is the target nucleus.
For a spin-zero nucleus and $T \ll E$,
where
$T$ is the nuclear kinetic recoil energy and $E$ is the neutrino energy,
the differential cross section is given by
\begin{align}
\dfrac{d\sigma_{\nu_{\ell}\text{-}\mathcal{N}}}{d T}
\null & \null
(E,T)
\simeq
\dfrac{G_{\text{F}}^2 M}{\pi}
\left(
1 - \dfrac{M T}{2 E^2}
\right)
\nonumber
\\
\null & \null
\times
\Big\{
\left[
\left( g_{V}^{p} - \tilde{Q}_{\ell\ell} \right)
Z
F_{Z}(|\vet{q}|^2)
+
g_{V}^{n}
N
F_{N}(|\vet{q}|^2)
\right]^2
\nonumber
\\
\null & \null
\hspace{1cm}
+
Z^2
F_{Z}^2(|\vet{q}|^2)
\sum_{\ell'\neq\ell}
|\tilde{Q}_{\ell'\ell}|^2
\Big\}
,
\label{cs}
\end{align}
with\footnote{We neglect the radiative corrections to $g_{V}^{p}$ and $g_{V}^{n}$
(see Ref.~\cite{Barranco:2005yy}),
that are too small to affect our results.}
\begin{align}
g_{V}^{p}
=
\null & \null
\dfrac{1}{2} - 2 \sin^2\!\vartheta_{W}
,
\label{gVp}
\\
g_{V}^{n}
=
\null & \null
- \dfrac{1}{2}
,
\label{gVn}
\end{align}
\begin{equation}
\tilde{Q}_{\ell\ell'}
=
\frac{2}{3} \, m_{W}^2 \sin^2\!\vartheta_{W} \langle{r}_{\nu_{\ell\ell'}}^2\rangle
.
\label{Qab}
\end{equation}
In Eq.~(\ref{cs}) one can distinguish the effects of the diagonal charge radii
$\langle{r}_{\nu_{\ell}}^2\rangle \equiv \langle{r}_{\nu_{\ell\ell}}^2\rangle$
that contribute through the addition of
$\tilde{Q}_{\ell\ell}$
to
$g_{V}^{p}$,
which is equivalent to the shift in Eq.~(\ref{shift}).
This contribution affects only the protons in the nucleus,
whose number is given by $Z$.
On the other hand,
the transition charge radii
$\langle{r}_{\nu_{\ell\ell'}}^2\rangle$
with $\ell\neq\ell'$
contribute to the cross section trough an additional term
proportional to $Z^2$.
The neutrons, whose number is given by $N$, of course do not interact with the neutrino charge radii.
Note also that the charge radii of antineutrinos are related to those of neutrinos by\footnote{
From Eqs.~(3.48) and (7.33) of Ref.~\cite{Giunti:2014ixa},
we have
$\langle{r}_{\bar\nu_{jk}}^2\rangle = - \langle{r}_{\nu_{kj}}^2\rangle$.
Since
$
\langle{r}_{\bar\nu_{\ell\ell'}}^2\rangle
=
\sum_{j,k} U_{\ell j} U_{\ell' k}^{*} \langle{r}_{\nu_{jk}}^2\rangle
$,
from Eq.~(\ref{basis})
we obtain
$
\langle{r}_{\bar\nu_{\ell\ell'}}^2\rangle
=
- \langle{r}_{\nu_{\ell'\ell}}^2\rangle
$.
}
\begin{equation}
\langle{r}_{\bar\nu_{\ell\ell'}}^2\rangle
=
- \langle{r}_{\nu_{\ell'\ell}}^2\rangle
,
\label{antinu}
\end{equation}
but also
$g_{V}^{p}$ and $g_{V}^{n}$
change sign for antineutrinos.
Therefore,
the diagonal charge radii of neutrinos and antineutrinos
generate the same shift of
$\sin^2\!\vartheta_{W}$
in Eq.~(\ref{shift}).

For the proton and neutron contributions in Eq.~(\ref{cs})
we take into account the corresponding nuclear form factors
$F_{Z}(|\vet{q}|^2)$
and
$F_{N}(|\vet{q}|^2)$,
which
are the Fourier transforms of the corresponding nucleon
distribution in the nucleus and
describe the loss of coherence for
$|\vet{q}| R \gtrsim 1$,
where $R$ is the nuclear radius.
These distributions are usually expressed with an
appropriate parameterization which depends on two parameters:
the rms radius $R$ and the surface thickness $s$.
The most common parameterizations are the Fermi,
symmetrized Fermi~\cite{Piekarewicz:2016vbn},
and Helm~\cite{Helm:1956zz}.
Since these different parameterization are practically equivalent in the
analysis of COHERENT data~\cite{Cadeddu:2017etk},
for simplicity in the following we use only the Helm parameterization~\cite{Helm:1956zz}
\begin{equation}
F(|\vet{q}|^2)
=
3
\,
\dfrac{j_{1}(|\vet{q}| R_{0})}{|\vet{q}| R_{0}}
\,
e^{- |\vet{q}|^2 s^2 / 2}
,
\label{ffHelm}
\end{equation}
where
$
j_{1}(x) = \sin(x) / x^2 - \cos(x) / x
$
is the spherical Bessel function of order one
and $R_{0}$ is related to the rms radius $R$ by
\begin{equation}
R^2 = \dfrac{3}{5} \, R_{0}^2 + 3 s^2
.
\label{RnHelm}
\end{equation}
For the surface thickness $s$ we consider the value $s = 0.9 \, \text{fm}$
that was determined for the proton form factor of similar nuclei
\cite{Friedrich:1982esq}.

\begin{table*}[t!]
\begin{center}
\begin{tabular}{c|cc|cc|cc|cc|}
\\
&
\multicolumn{4}{c|}{Spectrum}
&
\multicolumn{4}{c|}{Spectrum and time}
\\
&
\multicolumn{2}{c|}{$\langle{r}_{\nu_{e}}^2\rangle$ and $\langle{r}_{\nu_{\mu}}^2\rangle$ only}
&
\multicolumn{2}{c|}{All $\langle{r}_{\nu}^2\rangle$}
&
\multicolumn{2}{c|}{$\langle{r}_{\nu_{e}}^2\rangle$ and $\langle{r}_{\nu_{\mu}}^2\rangle$ only}
&
\multicolumn{2}{c|}{All $\langle{r}_{\nu}^2\rangle$}
\\
&
Fixed $R_{n}$
&
Free $R_{n}$
&
Fixed $R_{n}$
&
Free $R_{n}$
&
Fixed $R_{n}$
&
Free $R_{n}$
&
Fixed $R_{n}$
&
Free $R_{n}$
\\
\hline
$\chi^{2}_{\text{min}}$
&
2.6
&
2.6
&
2.6
&
2.6
&
153.9
&
153.7
&
154.0
&
153.8
\\
NDF
&
142
&
140
&
139
&
137
&
142
&
140
&
139
&
137
\\
GoF
&
100\%
&
100\%
&
100\%
&
100\%
&
23\%
&
20\%
&
18\%
&
15\%
\\
$\langle{r}_{\nu_{e}}^2\rangle$
&
$(
-76
,
27
)$
&
$(
-77
,
32
)$
&
$(
-74
,
27
)$
&
$(
-73
,
29
)$
&
$(
-56
,
6
)$
&
$(
-56
,
8
)$
&
$(
-56
,
6
)$
&
$(
-55
,
6
)$
\\
$\langle{r}_{\nu_{\mu}}^2\rangle$
&
$(
-63
,
14
)$
&
$(
-64
,
17
)$
&
$(
-62
,
13
)$
&
$(
-61
,
15
)$
&
$(
-61
,
10
)$
&
$(
-60
,
10
)$
&
$(
-60
,
9
)$
&
$(
-59
,
9
)$
\\
$|\langle{r}_{\nu_{e\mu}}^2\rangle|$
&
&
&
$< 31$
&
$< 31$
&
&
&
$< 28$
&
$< 28$
\\
$|\langle{r}_{\nu_{e\tau}}^2\rangle|$
&
&
&
$< 51$
&
$< 51$
&
&
&
$< 31$
&
$< 30$
\\
$|\langle{r}_{\nu_{\mu\tau}}^2\rangle|$
&
&
&
$< 38$
&
$< 38$
&
&
&
$< 34$
&
$< 35$
\end{tabular}
\caption{ \label{tab:fit}
Results of the fits of the COHERENT data.
The limits on the charge radii are at 90\% CL and in units of
$10^{-32} \, \text{cm}^2$.
}
\end{center}
\end{table*}

The COHERENT experiment measured
the coherent elastic scattering of neutrinos on
$^{133}\text{Cs}$
and
$^{127}\text{I}$.
Hence,
the total cross section is given by
\begin{equation}
\dfrac{d\sigma_{\nu\text{-}\text{CsI}}}{d T}
=
\dfrac{d\sigma_{\nu\text{-}\text{Cs}}}{d T}
+
\dfrac{d\sigma_{\nu\text{-}\text{I}}}{d T}
,
\label{cs02}
\end{equation}
with
$N_{\text{Cs}} = 78$,
$Z_{\text{Cs}} = 55$,
$N_{\text{I}} = 74$, and
$Z_{\text{I}} = 53$.
We neglect the small axial contribution due to the unpaired valence protons~\cite{Barranco:2005yy}.

In our analysis of the COHERENT data we use the values of the
rms radii of the proton distribution of
$^{133}\text{Cs}$ and $^{127}\text{I}$
that have been determined with high accuracy with
muonic atom spectroscopy~\cite{Fricke:1995zz}:
\begin{align}
\null & \null
R_{p}({}^{133}\text{Cs}) = 4.804 \, \text{fm}
,
\label{RpCs}
\\
\null & \null
R_{p}({}^{127}\text{I}) = 4.749 \, \text{fm}
.
\label{RpI}
\end{align}
On the other hand, there is no experimental determination of the
value of the rms radii of the neutron distribution of
$^{133}\text{Cs}$ and $^{127}\text{I}$,
except that obtained in Ref.~\cite{Cadeddu:2017etk}
from the analysis of the COHERENT data assuming the absence of
effects due to the neutrino charge radii and non-standard interactions.
Therefore,
in order to extract information on the neutrino charge radii from the COHERENT data
we adopt the following two approaches:

\begin{figure*}[!t]
\centering
\setlength{\tabcolsep}{0pt}
\begin{tabular}{cc}
\subfigure[]{\label{fig:dim-2-4}
\includegraphics*[width=0.49\linewidth]{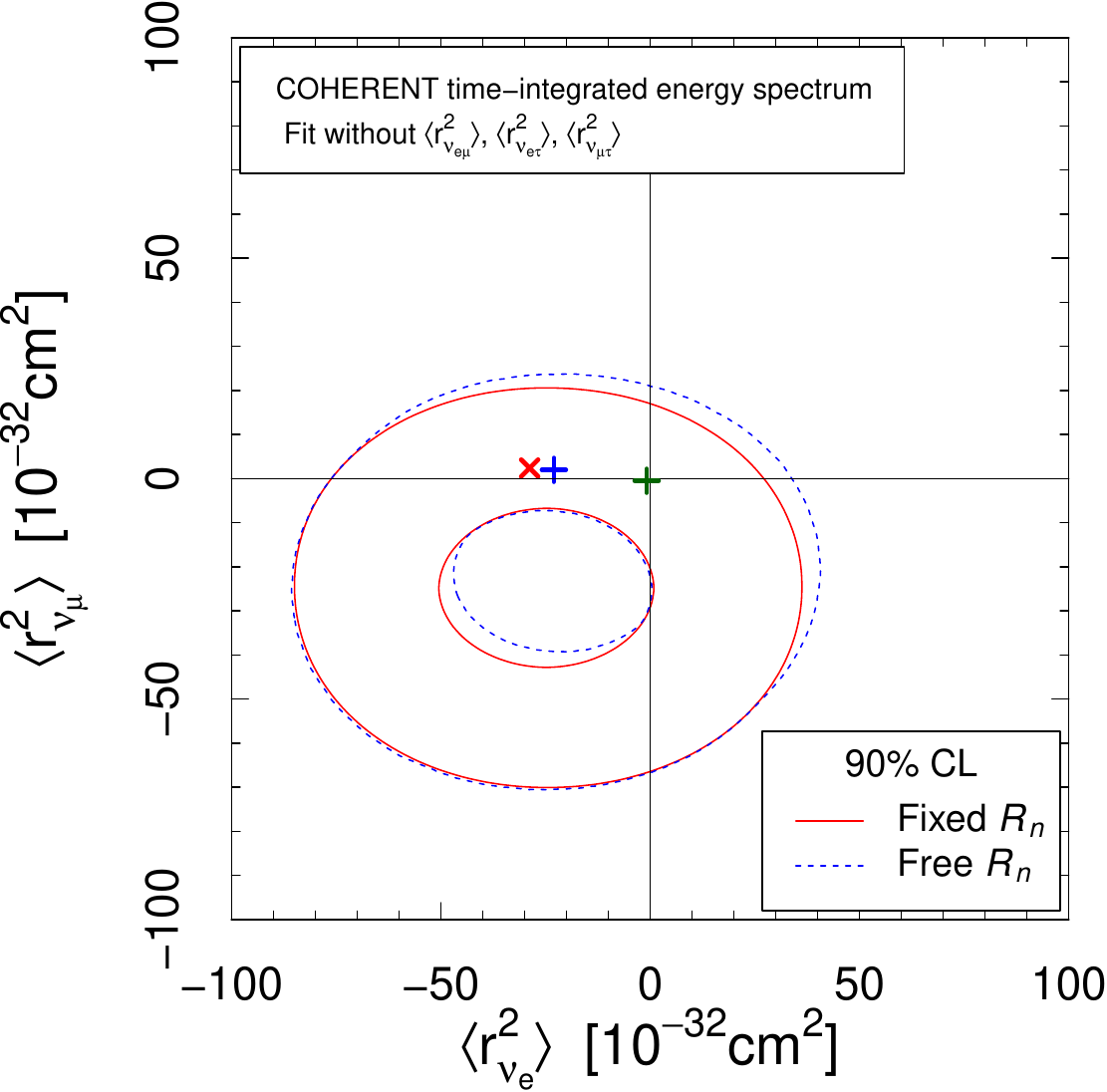}
}
&
\subfigure[]{\label{fig:dim-5-7}
\includegraphics*[width=0.49\linewidth]{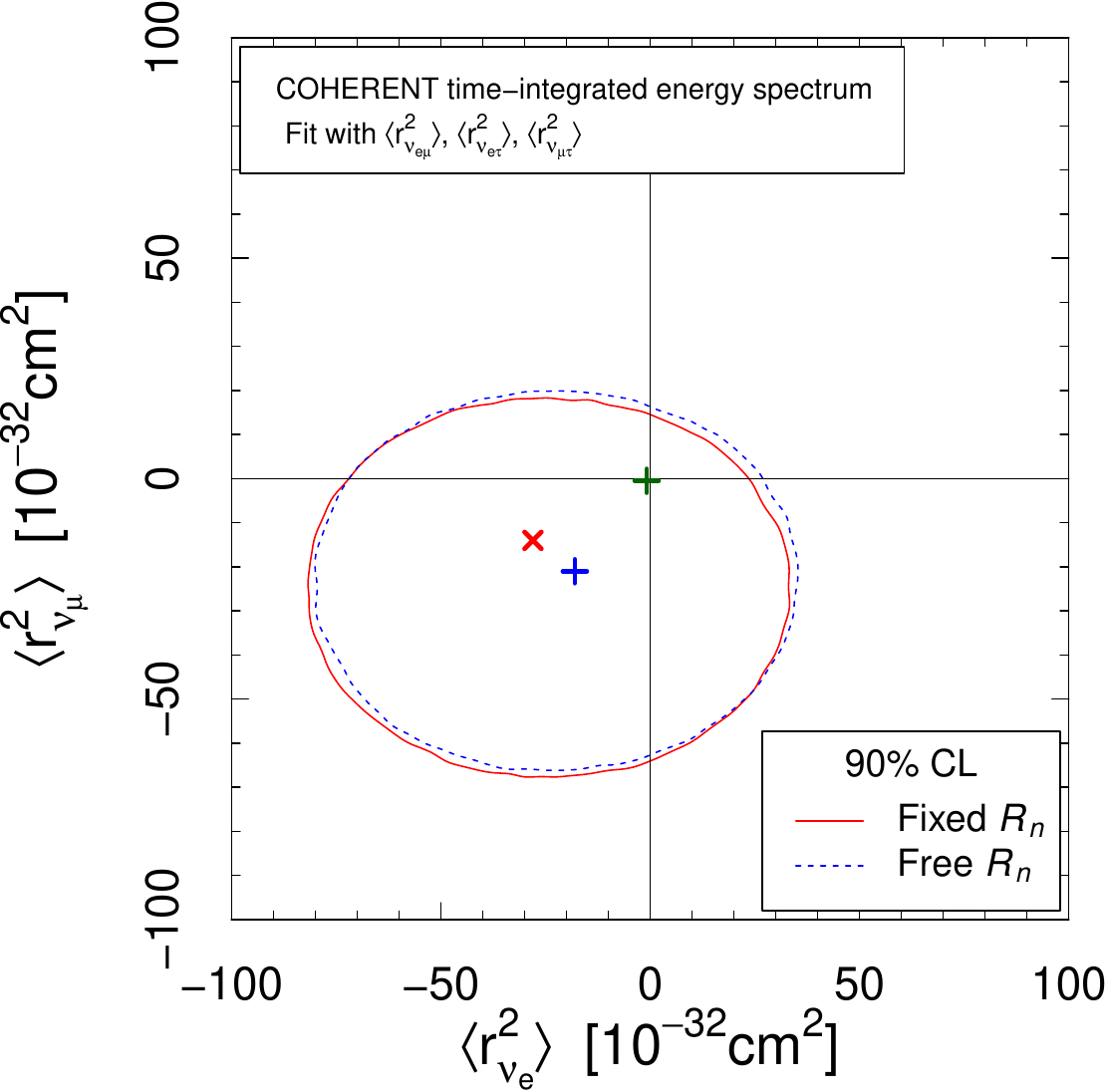}
}
\end{tabular}
\caption{ \label{fig:dim}
90\% CL allowed regions in the
$\langle{r}_{\nu_{e}}^2\rangle$--$\langle{r}_{\nu_{\mu}}^2\rangle$
plane
obtained from the fit of the
time-integrated COHERENT energy spectrum
without \subref{fig:dim-2-4}
and
with \subref{fig:dim-5-7}
the transition charge radii.
The red and blue points indicate the best-fit values.
The green point near the origin indicates the Standard Model values
in Eqs.~(\ref{reSM}) and (\ref{rmSM}).
}
\end{figure*}

\begin{description}

\item[\textbf{Fixed $\mathbf{R_{n}}$}]
We assume the theoretical values
\begin{align}
\null & \null
R_{n}({}^{133}\text{Cs}) = 5.01 \, \text{fm}
,
\label{RnCs}
\\
\null & \null
R_{n}({}^{127}\text{I}) = 4.94 \, \text{fm}
,
\label{RnI}
\end{align}
obtained in the relativistic mean field (RMF) NL-Z2 \cite{Bender:1999yt}
nuclear model calculated in Ref.~\cite{Cadeddu:2017etk}.
This is a reasonable assumption taking into account that the values of the
rms radii of the proton distribution of
$^{133}\text{Cs}$ and $^{127}\text{I}$
calculated with the RMF NL-Z2,
$R_{p}({}^{133}\text{Cs}) = 4.79 \, \text{fm}$
and
$R_{p}({}^{127}\text{I}) = 4.73 \, \text{fm}$,
are in approximate agreement with the experimental values in Eqs.~(\ref{RpCs}) and (\ref{RpI}).

\item[\textbf{Free $\mathbf{R_{n}}$}]
We perform a fit of the coherent data with free
$R_{n}({}^{133}\text{Cs})$
and
$R_{n}({}^{127}\text{I})$
which are allowed to vary in a suitable interval.
For the lower bounds of the allowed ranges of
$R_{n}({}^{133}\text{Cs})$
and
$R_{n}({}^{127}\text{I})$
we took the corresponding experimental values of
$R_{p}({}^{133}\text{Cs})$
and
$R_{p}({}^{127}\text{I})$
in Eqs.~(\ref{RpCs}) and (\ref{RpI}).
These are very reliable lower bounds,
because in $^{133}\text{Cs}$ and $^{127}\text{I}$
there are about 20 more neutrons than protons and all model calculations predict
$R_{n}>R_{p}$.
The upper bounds for
$R_{n}({}^{133}\text{Cs})$
and
$R_{n}({}^{127}\text{I})$
are more arbitrary, since there is no experimental information.
However,
the parity-violating PREX experiment measured
$
R_{n}(^{208}\text{Pb})
=
5.75 \pm 0.18
\,
\text{fm}
$
\cite{Abrahamyan:2012gp,Horowitz:2012tj}.
Since it is very unlikely that
$R_{n}({}^{133}\text{Cs})$
and
$R_{n}({}^{127}\text{I})$
are larger than
$R_{n}(^{208}\text{Pb})$,
we adopt the upper bound of 6 fm.
In any case, we have checked that the limits that we obtain for the neutrino charge radii
are stable within reasonable changes of the upper bounds for the allowed ranges of
$R_{n}({}^{133}\text{Cs})$
and
$R_{n}({}^{127}\text{I})$.

\end{description}

\section{Fit of the time-integrated COHERENT energy spectrum}
\label{sec:spectrum}

As a first step,
we fitted the time-integrated COHERENT energy spectrum
with the same method as in Ref.~\cite{Cadeddu:2017etk},
using the precise information in the
COHERENT data release~\cite{Akimov:2018vzs}.
We considered the least-squares function
\begin{align}
\chi^2
=
\null & \null
\sum_{i=4}^{15}
\left(
\dfrac{
N_{i}^{\text{exp}}
-
\left(1+\alpha\right) N_{i}^{\text{th}}
-
\left(1+\beta\right) B_{i}
}{ \sigma_{i} }
\right)^2
\nonumber
\\
\null & \null
+
\left( \dfrac{\alpha}{\sigma_{\alpha}} \right)^2
+
\left( \dfrac{\beta}{\sigma_{\beta}} \right)^2
.
\label{chi}
\end{align}
For each energy bin $i$,
$N_{i}^{\text{exp}}$
and
$N_{i}^{\text{th}}$
are, respectively,
the experimental and theoretical number of events,
$B_{i}$ is the estimated number of background events, and
$\sigma_{i}$ is the statistical uncertainty.
$\alpha$ and $\beta$
are nuisance parameters which quantify,
respectively,
the systematic uncertainty of the signal rate
and
the systematic uncertainty of the background rate
with
corresponding standard deviations
$\sigma_{\alpha} = 0.28$
and
$\sigma_{\beta} = 0.25$
\cite{Akimov:2017ade}.
The fit is restricted to the bin numbers from 4 to 15
where the acceptance function is non-zero
and
the linear relation
$
N_{\text{PE}}
=
1.17 \, T / \text{keV}
$
between the observed number of photoelectrons $N_{\text{PE}}$
and the nuclear kinetic recoil energy $T$
is reliable~\cite{Akimov:2017ade}.

\begin{figure*}[t!]
\begin{center}
\includegraphics*[width=\linewidth]{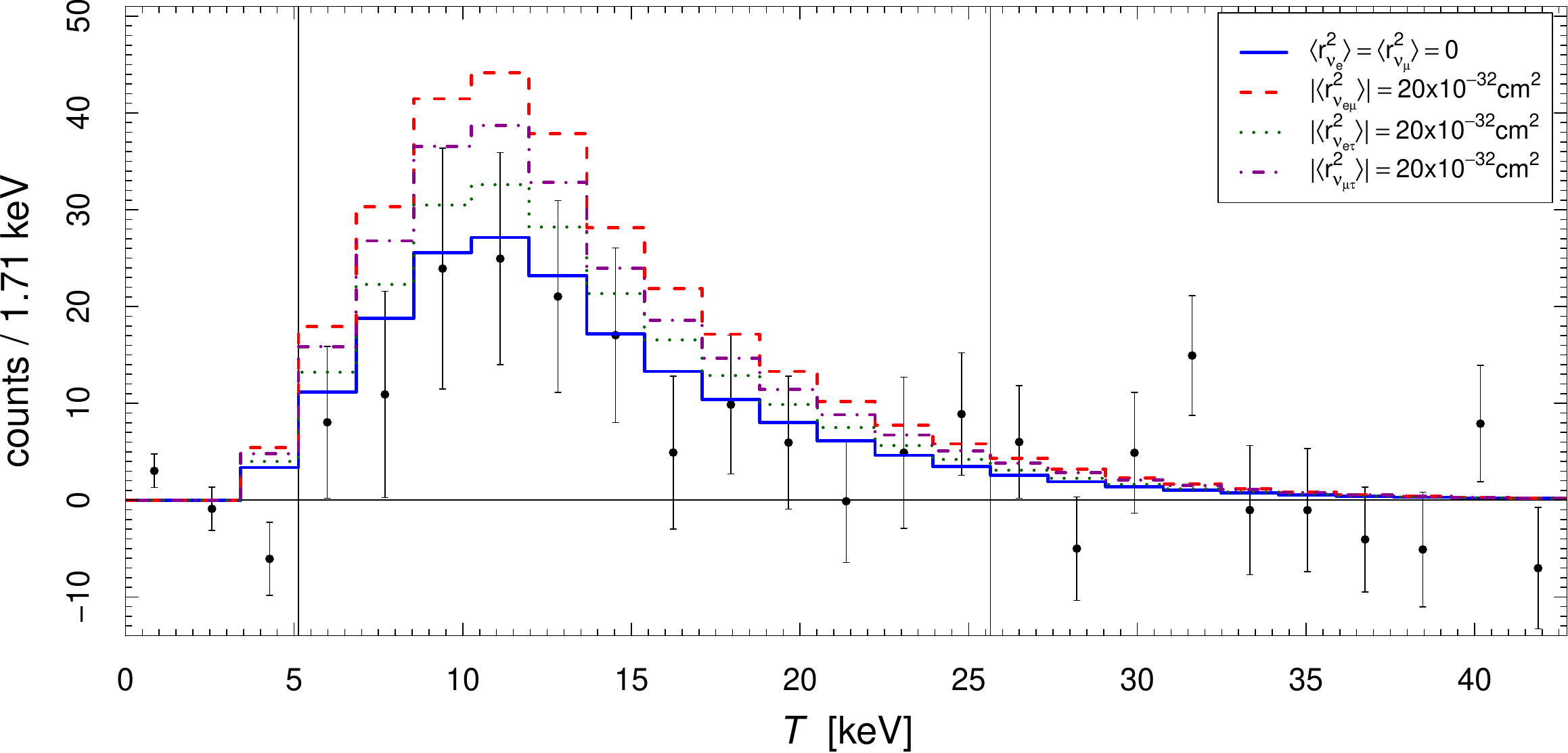}
\end{center}
\caption{ \label{fig:histTofd}
COHERENT data \cite{Akimov:2017ade,Akimov:2018vzs} versus the nuclear kinetic recoil energy $T$.
The blue histogram shows the theoretical predictions
without neutrino charge radii and the theoretical values
of the rms radii of the neutron distributions in Eqs.~(\ref{RnCs}) and (\ref{RnI}).
The dashed red, dotted green and dash-dotted magenta histograms show the theoretical predictions obtained
by adding the contributions of the neutrino transition charge radii reported in the legend.
The two vertical lines at
5.13 and 25.64 keV indicate the range of our analysis, as explained after Eq.~(\ref{chi}).
}
\end{figure*}

The theoretical number of coherent elastic scattering events
$N_{i}^{\text{th}}$
in each energy bin $i$
depends on the neutrino charge radii and on the nuclear form factors.
It is given by
\begin{equation}
N_{i}^{\text{th}}
=
N_{\text{CsI}}
\int_{T_{i}}^{T_{i+1}} d T
\int_{E_{\text{min}}} d E
\,
A(T)
\,
\frac{d N_{\nu}}{d E}
\,
\dfrac{d\sigma_{\nu\text{-}\text{CsI}}}{d T}
,
\label{Nth}
\end{equation}
where $N_{\text{CsI}}$
is the number of CsI
in the detector
(given by
$ N_{\text{A}} M_{\text{det}} / M_{\text{CsI}}$,
where
$ N_{\text{A}} $
is the Avogadro number,
$ M_{\text{det}} = 14.6 \,\text{kg} $,
is the detector mass, and
$ M_{\text{CsI}} = 259.8 $
is the molar mass of CsI),
$ E_{\text{min}} = \sqrt{M T / 2} $,
$A(T)$
is the acceptance function
and
$d N_{\nu} / d E$
is the neutrino flux integrated over the experiment lifetime.
Neutrinos at the Spallation Neutron Source consist of a prompt component of monochromatic
$\nu_\mu$ from stopped pion decays,
$\pi^+\to \mu^++\nu_\mu$,
and two delayed components of
$\bar{\nu}_\mu$ and $\nu_{e}$
from the subsequent muon decays, $\mu^+\to e^+ + \bar{\nu}_\mu + \nu_{e}$.
The total flux $d N_{\nu} / d E$ is the sum of
\begin{align}
\frac{d N_{\nu_{\mu}}}{d E}
=
\null & \null
\eta
\,
\delta\!\left(
E - \dfrac{ m_{\pi}^2 - m_{\mu}^2 }{ 2 m_{\pi} }
\right)
,
\label{numu}
\\
\frac{d N_{\nu_{\bar\mu}}}{d E}
=
\null & \null
\eta
\,
\dfrac{ 64 E^2 }{ m_{\mu}^3 }
\left(
\dfrac{3}{4} - \dfrac{E}{m_{\mu}}
\right)
,
\label{numubar}
\\
\frac{d N_{\nu_{e}}}{d E}
=
\null & \null
\eta
\,
\dfrac{ 192 E^2 }{ m_{\mu}^3 }
\left(
\dfrac{1}{2} - \dfrac{E}{m_{\mu}}
\right)
,
\label{nue}
\end{align}
for
$E \leq m_{\mu} / 2 \simeq 52.8 \, \text{MeV}$,
with the normalization factor
$ \eta = r N_{\text{POT}} / 4 \pi L^2 $,
where
$r=0.08$ is the number of neutrinos per flavor
that are produced for each proton on target,
$ N_{\text{POT}} = 1.76\times 10^{23} $
is the number of proton on target
and $ L = 19.3 \, \text{m} $
is the distance between the source and the COHERENT CsI detector~\cite{Akimov:2017ade}.
The three neutrino spectra are illustrated in Fig.~\ref{fig:spec}.
Note that for the $\bar\nu_{\mu}$ spectrum it is important to take into account the relation
in Eq.~(\ref{antinu}).

Figure~\ref{fig:histT0}
illustrates the effects of the neutrino charge radii
$\langle{r}_{\nu_{e}}^2\rangle$
and
$\langle{r}_{\nu_{\mu}}^2\rangle$
in the fit of the COHERENT data.
Note that for
$|\langle{r}_{\nu_{e,\mu}}^2\rangle| \simeq 20 \times 10^{-32} \, \text{cm}^2$
the contribution of the proton term in Eq.~(\ref{cs})
becomes similar to that of the neutron term,
which is dominant in the absence of neutrino charge radii.
Therefore, since the uncertainties of the COHERENT data are quite large,
we expect to obtain limits on the neutrino charge radii
of the order of
$20 \times 10^{-32} \, \text{cm}^2$.
Indeed,
one can see from Fig.~\ref{fig:histT0}
that values of
$\langle{r}_{\nu_{e}}^2\rangle$
and
$\langle{r}_{\nu_{\mu}}^2\rangle$
of this size generate histograms that fit badly the data.

The first two columns in Tab.~\ref{tab:fit}
and Fig.~\ref{fig:dim-2-4}
show the results of the fit of the
time-integrated COHERENT energy spectrum
considering only the effects of the diagonal neutrino charge radii
$\langle{r}_{\nu_{e}}^2\rangle$ and $\langle{r}_{\nu_{\mu}}^2\rangle$.
This restriction is appropriate for the measurement
of the neutrino charge radii
predicted by the Standard Model
[Eqs.~(\ref{reSM}) and (\ref{rmSM})],
where there are no transition charge radii.
From Tab.~\ref{tab:fit}
one can see that the fits of the data are very good
both with fixed and free $R_{n}$.
The allowed ranges of
$\langle{r}_{\nu_{e}}^2\rangle$ and $\langle{r}_{\nu_{\mu}}^2\rangle$
are some tens of $10^{-32} \, \text{cm}^2$,
as expected.
From Tab.~\ref{tab:fit} and Fig.~\ref{fig:dim-2-4}
one can see that the allowed ranges of
$\langle{r}_{\nu_{e}}^2\rangle$ and $\langle{r}_{\nu_{\mu}}^2\rangle$
are almost insensitive to the value of $R_{n}$.

Let us now consider the fit of the
time-integrated COHERENT energy spectrum
in the complete theory including the effects of possible
neutrino transition charge radii.
The third and fourth columns in Tab.~\ref{tab:fit}
and Fig.~\ref{fig:dim-5-7}
show the results of the fits with fixed and free $R_{n}$.
The marginal allowed ranges of
$\langle{r}_{\nu_{e}}^2\rangle$ and $\langle{r}_{\nu_{\mu}}^2\rangle$
do not change significantly with respect to those
obtained without the transition charge radii,
but the allowed region in the
$\langle{r}_{\nu_{e}}^2\rangle$--$\langle{r}_{\nu_{\mu}}^2\rangle$
plane becomes simply connected.
The hole in the allowed region with the diagonal charge radii alone
is due to a cancellation between the
weak interaction and charge radii contributions to the cross section.
The hole disappears when the neutrino transition charge radii are taken into account,
because their contribution to the cross section can fit the data.

It is interesting that we obtained for the first time
constraints on the neutrino transition charge radii.
Their effect is illustrated in Fig.~\ref{fig:histTofd},
where one can see that they always increase the predicted event rate,
because their contribution adds incoherently to weak interactions
in the cross section (\ref{cs}).
From Tab.~\ref{tab:fit} and Fig.~\ref{fig:dim-5-7}
one can also see that the limits on
$\langle{r}_{\nu_{e}}^2\rangle$ and $\langle{r}_{\nu_{\mu}}^2\rangle$
are not sensitive to the assumed value of $R_{n}$,
because the transition charge radii can compensate the effects of the variations of $R_{n}$.

Let us also note that the best-fit values of
$\langle{r}_{\nu_{e}}^2\rangle$--$\langle{r}_{\nu_{\mu}}^2\rangle$
shown by points in Fig.~\ref{fig:dim}
correspond to a large negative value of
$\langle{r}_{\nu_{e}}^2\rangle$
in the left panel obtained without the transition charge radii,
and
large negative values of both
$\langle{r}_{\nu_{e}}^2\rangle$
and
$\langle{r}_{\nu_{\mu}}^2\rangle$
in the right panel obtained with the transition charge radii.
However,
these indications do not have a sufficient statistical significance
and
it is wise to rely only on the 90\% CL contours in Fig.~\ref{fig:dim},
which include the Standard Model values in Eqs.~(\ref{reSM}) and (\ref{rmSM}).

\begin{figure}[t!]
\begin{center}
\includegraphics*[width=\linewidth]{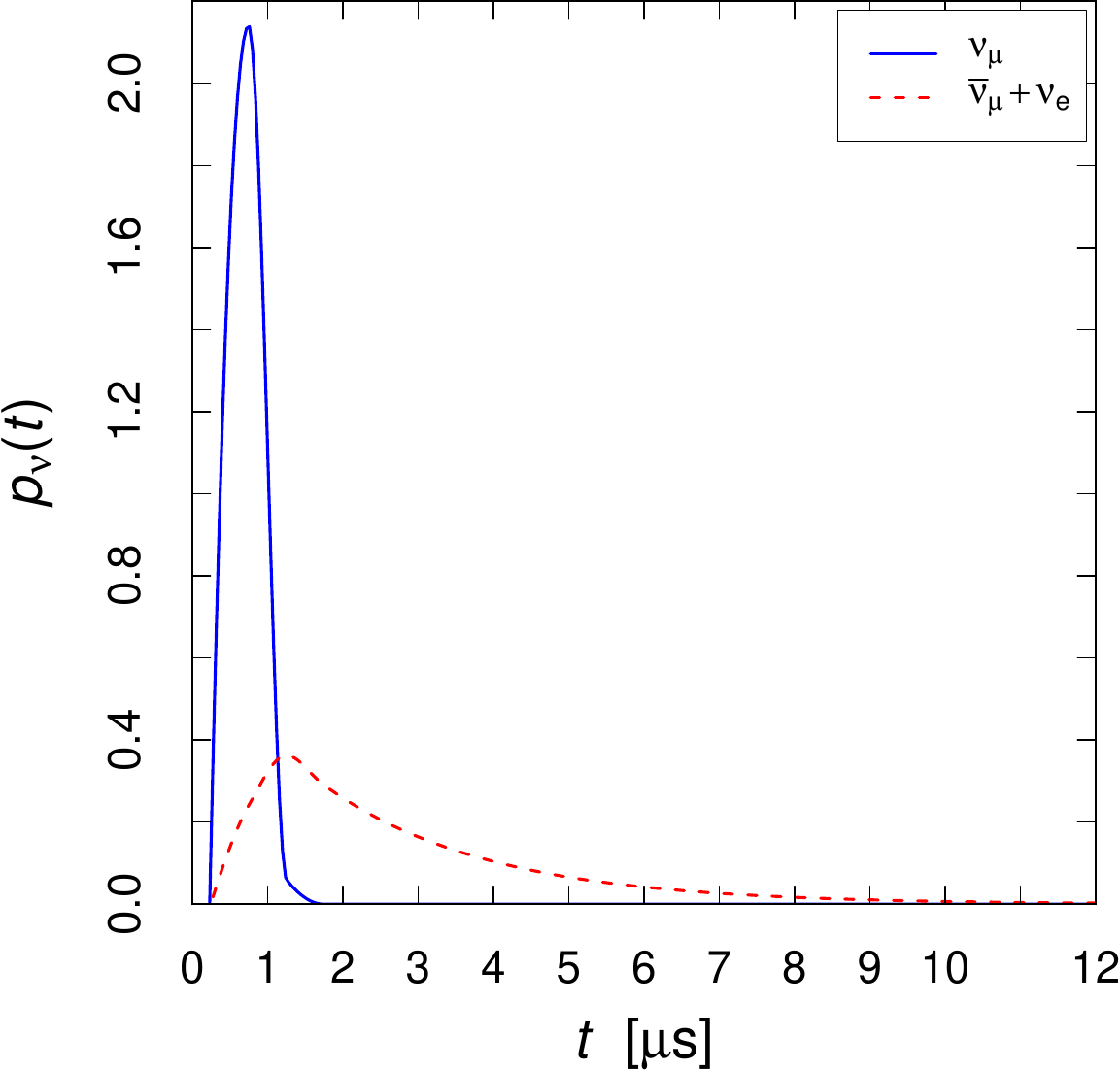}
\end{center}
\caption{ \label{fig:timpdf}
The time distribution of arrivals after protons-on-targets
of
prompt $\nu_{\mu}$'s
and delayed
$\bar\nu_{\mu}$'s and $\nu_{e}$'s
in the COHERENT experiment~\cite{Akimov:2017ade,Akimov:2018vzs}.
}
\end{figure}

\begin{figure*}[!t]
\centering
\setlength{\tabcolsep}{0pt}
\begin{tabular}{cc}
\subfigure[]{\label{fig:sft-2-4}
\includegraphics*[width=0.49\linewidth]{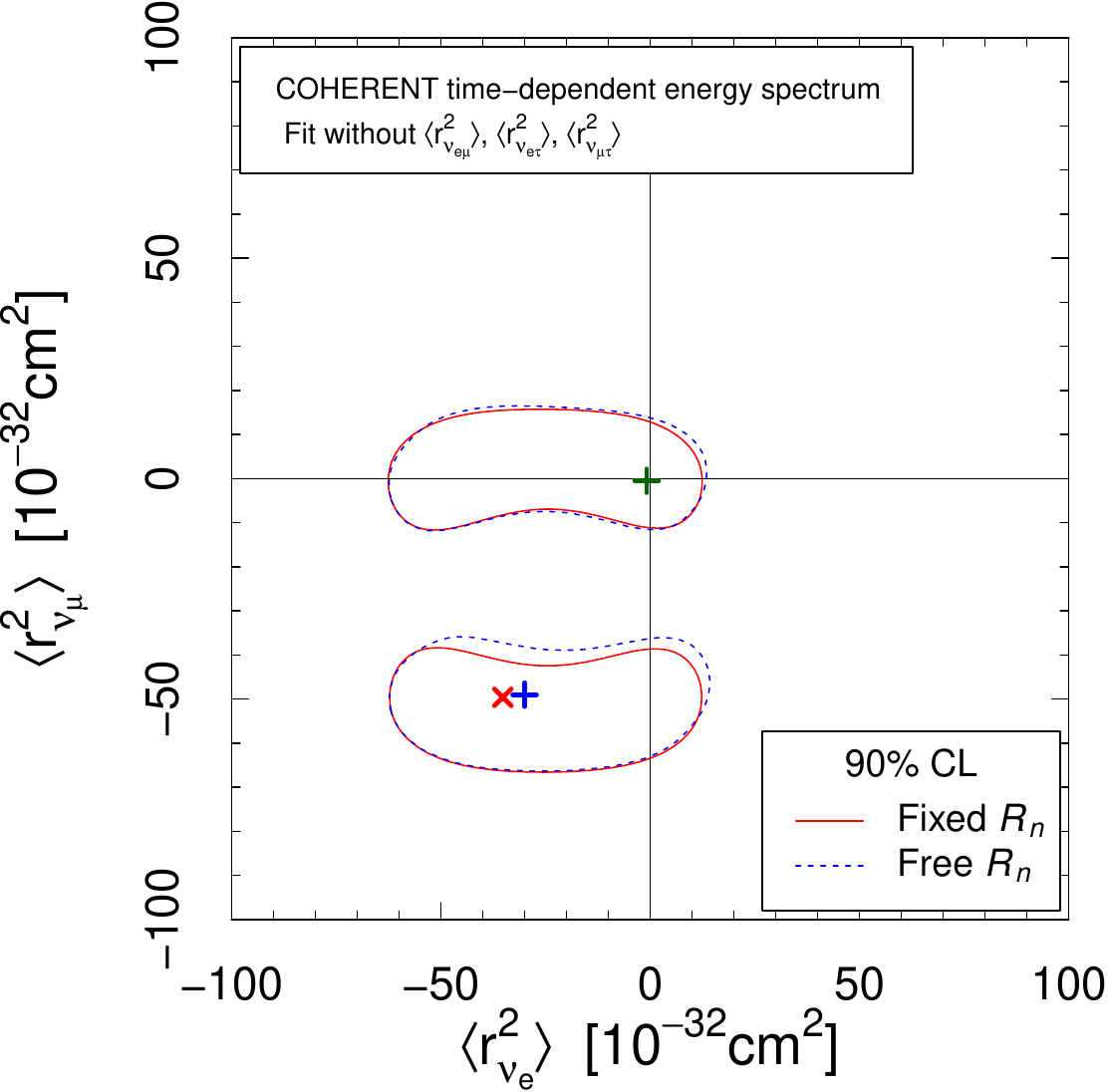}
}
&
\subfigure[]{\label{fig:sft-5-7}
\includegraphics*[width=0.49\linewidth]{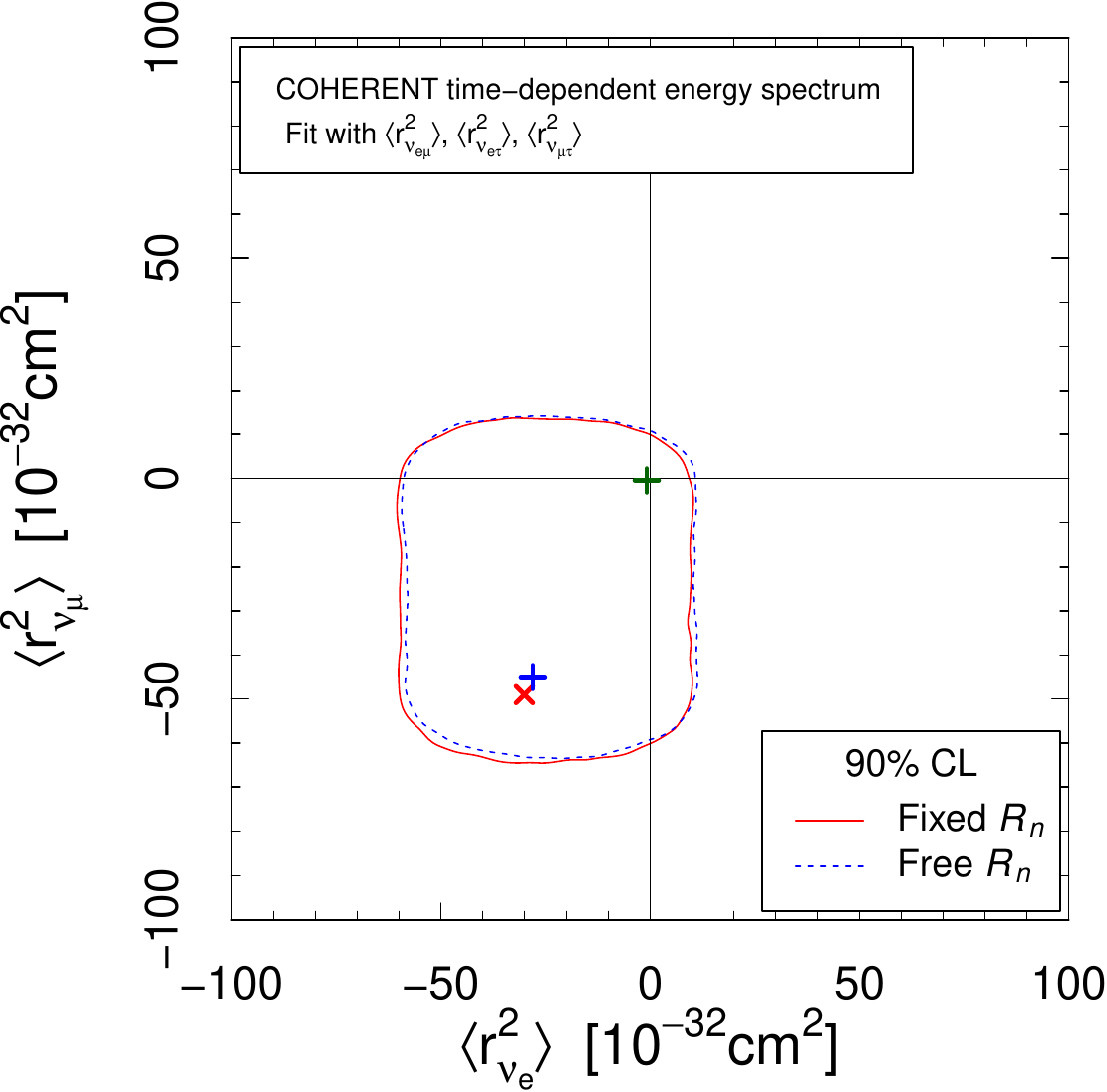}
}
\end{tabular}
\caption{ \label{fig:sft}
90\% CL allowed regions in the
$\langle{r}_{\nu_{e}}^2\rangle$--$\langle{r}_{\nu_{\mu}}^2\rangle$
plane
obtained from the fit of the
time-dependent COHERENT energy spectrum
without \subref{fig:sft-2-4}
and
with \subref{fig:sft-5-7}
the transition charge radii.
The red and blue points indicate the best-fit values.
The green point near the origin indicates the Standard Model values
in Eqs.~(\ref{reSM}) and (\ref{rmSM}).
}
\end{figure*}

\section{Fit of the time-dependent COHERENT data}
\label{sec:time}

The data release in Ref.~\cite{Akimov:2018vzs} of the COHERENT collaboration
contains the complete information on the energy and arrival time of the observed events.
The time of arrival after protons-on-targets is an important information for the study of neutrino properties,
because $\nu_{\mu}$'s produced by $\pi^{+}$ decay arrive promptly within about
$1.5 \, \mu\text{s}$,
whereas $\bar\nu_{\mu}$'s and $\nu_e$'s produced by $\mu^{+}$ arrive in a time interval of about
$10 \, \mu\text{s}$
(see Fig.~\ref{fig:timpdf}).
Hence,
the time distribution of the data increases the information on the
difference between the properties of
$\nu_\mu$
and those of
$\bar{\nu}_\mu$ and $\nu_e$.

We analyzed the time-dependent COHERENT data in the following way,
according to the prescriptions given in the COHERENT data release~\cite{Akimov:2018vzs}.
Since in the binning of the events in energy and time there are some bins with zero count,
we considered the Poissonian least-squares function~\cite{Baker:1983tu}
\begin{align}
\chi^2
=
\null & \null
2
\sum_{i=4}^{15}
\sum_{j=1}^{12}
\Bigg[
\left( 1 + \alpha \right) N_{ij}^{\text{th}}
+
\left( 1 + \beta \right) B_{ij}
\nonumber
\\
\null & \null
\hspace{1cm}
+
\left( 1 + \gamma \right) N_{ij}^{\text{bck}}
-
N_{ij}^{\text{C}}
\nonumber
\\
\null & \null
\hspace{-1cm}
+
N_{ij}^{\text{C}}
\ln\!\left(
\frac{ N_{ij}^{\text{C}} }{
\left( 1 + \alpha \right) N_{ij}^{\text{th}}
+
\left( 1 + \beta \right) B_{ij}
+
\left( 1 + \gamma \right) N_{ij}^{\text{bck}}
}
\right)
\Bigg]
\nonumber
\\
\null & \null
+
\left( \frac{\alpha}{\sigma_{\alpha}} \right)^2
+
\left( \frac{\beta}{\sigma_{\beta}} \right)^2
+
\left( \frac{\gamma}{\sigma_{\gamma}} \right)^2
,
\label{chi2time}
\end{align}
where
$i$ is the index of the energy bins,
$j$ is the index of the time bins,
$N_{ij}^{\text{th}}$ are the theoretical predictions that depend on the neutrino charge radii,
$N_{ij}^{\text{C}}$ are the coincidence (C) data, which contain signal and background events,
$B_{ij}$ are the estimated neutron-induced backgrounds,
$N_{ij}^{\text{bck}}$ are the estimated backgrounds obtained from the anti-coincidence (AC) data,
$\sigma_{\alpha} = 0.28$ is the systematic uncertainty of the signal rate,
$\sigma_{\beta} = 0.25$ is the systematic uncertainty of the neutron-induced background rate, and
$\sigma_{\gamma} = 0.05$ is the systematic uncertainty of the background estimated from the AC data
\cite{Akimov:2017ade,Akimov:2018vzs}.

Table~\ref{tab:fit}
and Fig.~\ref{fig:sft}
show the results of the fit of the time-dependent COHERENT data
with fixed and free $R_{n}$
and
without and with the neutrino transition charge radii.
Comparing these results
with the corresponding allowed intervals
in Tab.~\ref{tab:fit}
and the corresponding allowed regions in
Fig.~\ref{fig:dim}
obtained from the fit of the time-integrated COHERENT data,
one can see that, as expected,
the time information allows us to obtain better constraints on the neutrino charge radii.

The lower bound that we have obtained on
$\langle{r}_{\nu_{e}}^2\rangle$
is about one order of magnitude less stringent than the experimental
lower bounds in Tab.~\ref{tab:limits},
but the upper bound is comparable and confirms the results of those experiments.

One can also notice that the results on the neutrino charge radii are stable
under variations of $R_{n}$,
because the time dependence of the data is independent from
the rms radii of the neutron distributions
$^{133}\text{Cs}$
and
$^{127}\text{I}$.
Therefore,
the limits obtained for the neutrino charge radii are independent from the nuclear model.

Moreover,
from Tab~\ref{tab:fit} and
comparing Fig.~\ref{fig:dim-2-4} and \ref{fig:dim-5-7},
one can see that the inclusion in the analysis of the neutrino transition charge radii
has little effect on the determination of
$\langle{r}_{\nu_{e}}^2\rangle$
and
$\langle{r}_{\nu_{\mu}}^2\rangle$.
However,
let us notice that the analysis of the time-dependent COHERENT data
allows us to restrict the upper bounds on
the neutrino transition charge radii
obtained from the analysis of the time-integrated COHERENT data.

Let us finally note that,
as in the case of the fit of the time-integrated COHERENT data
commented at the end of Section~\ref{sec:spectrum},
the best-fit values of
$\langle{r}_{\nu_{e}}^2\rangle$ and $\langle{r}_{\nu_{\mu}}^2\rangle$
shown by points in Fig.~\ref{fig:sft}
correspond to large negative values of
$\langle{r}_{\nu_{e}}^2\rangle$
and
$\langle{r}_{\nu_{\mu}}^2\rangle$,
but
these indications do not have a sufficient statistical significance
and
we assert only the 90\% CL contours in Fig.~\ref{fig:sft},
which include the Standard Model values in Eqs.~(\ref{reSM}) and (\ref{rmSM}).

\section{Conclusions}
\label{sec:conclusions}

Coherent elastic neutrino-nucleus scattering
is a powerful tool to study neutrino and nuclear physics.
In this paper we have analyzed the first data on coherent elastic neutrino-nucleus scattering
obtained in the COHERENT experiment~\cite{Akimov:2017ade,Akimov:2018vzs}
in order to get information on the neutrino charge radii,
which are predicted in the Standard Model.
We obtained limits on the diagonal charge radii
$\langle{r}_{\nu_{e}}^2\rangle$
and
$\langle{r}_{\nu_{\mu}}^2\rangle$,
and on the transition charge radii
$\langle{r}_{\nu_{e\mu}}^2\rangle$,
$\langle{r}_{\nu_{e\tau}}^2\rangle$, and
$\langle{r}_{\nu_{\mu\tau}}^2\rangle$
from different analyses of the
time-integrated COHERENT energy spectrum
and
the time-dependent COHERENT data
taking into account the uncertainty of the neutron distributions in the
$^{133}\text{Cs}$
and
$^{127}\text{I}$
parameterized by the corresponding rms radii.

We have shown that the time information of the COHERENT data
allows us to restrict the allowed ranges of the charge radii.

We have obtained for the first time
limits on the neutrino transition charge radii
from experimental data (see Tab.~\ref{tab:fit}),
taking into account their effect in the cross section
according to Ref.~\cite{Kouzakov:2017hbc}:
\begin{equation}
\left(
|\langle{r}_{\nu_{e\mu}}^2\rangle|,
|\langle{r}_{\nu_{e\tau}}^2\rangle|,
|\langle{r}_{\nu_{\mu\tau}}^2\rangle|
\right)
<
\left(
28,
30,
35
\right)
\times 10^{-32}
\, \text{cm}^2
,
\label{rtrans}
\end{equation}
at 90\% CL,
marginalizing over reliable allowed intervals of the
rms radii of the neutron distributions of
$^{133}\text{Cs}$
and
$^{127}\text{I}$.
This is an interesting information on the physics beyond the Standard Model
which can generate the neutrino transition charge radii~\cite{NovalesSanchez:2008tn}.

The limits on the diagonal neutrino charge radii
$\langle{r}_{\nu_{e}}^2\rangle$
and
$\langle{r}_{\nu_{\mu}}^2\rangle$
that we have obtained are not better than the previous limits
in Tab.~\ref{tab:limits},
but our analysis confirms those limits
and hints at the likeliness of obtaining more stringent limits
with the oncoming more precise data of the COHERENT experiment~\cite{Akimov:2018ghi}
and other coherent elastic neutrino-nucleus scattering
experiments
(CONUS~\cite{Maneschg-Neutrino2018},
CONNIE~\cite{Aguilar-Arevalo:2016khx},
MINER~\cite{Agnolet:2016zir},
$\nu$-cleus~\cite{Strauss:2017cuu},
TEXONO~\cite{Singh:2017jow},
and others).

\begin{acknowledgments}
We would like to thank
Grayson Rich and Kate Scholberg
for enlightening discussions on the COHERENT experiment.
The work of Y.F. Li and Y.Y. Zhang is supported
in part by the National Natural Science Foundation of China under Grant No.~11835013, by the Strategic Priority Research Program of the Chinese Academy of Sciences under Grant No. XDA10010100.
The work of K.A. Kouzakov, Y.F. Li, Y.Y. Zhang, and A.I. Studenikin was supported in part by the joint project of
the National Natural Science Foundation of China under
Grant No. 1161101153, and the Russian Foundation for
Basic Research under Grant No.~17-52-53133~GFEN\_a.
Y.F. Li is also grateful for the support by the CAS Center for Excellence in Particle Physics (CCEPP).
K.A. Kouzakov  and  A.I. Studenikin are grateful for the
support by the Russian Foundation for
Basic Research under Grant No.~16-02-01023.
\end{acknowledgments}

%

\begin{thebibliography}{74}%
\makeatletter
\providecommand \@ifxundefined [1]{%
\@ifx{#1\undefined}
}%
\providecommand \@ifnum [1]{%
\ifnum #1\expandafter \@firstoftwo
\else \expandafter \@secondoftwo
\fi
}%
\providecommand \@ifx [1]{%
\ifx #1\expandafter \@firstoftwo
\else \expandafter \@secondoftwo
\fi
}%
\providecommand \natexlab [1]{#1}%
\providecommand \enquote [1]{``#1''}%
\providecommand \bibnamefont [1]{#1}%
\providecommand \bibfnamefont [1]{#1}%
\providecommand \citenamefont [1]{#1}%
\providecommand \href@noop [0]{\@secondoftwo}%
\providecommand \href [0]{\begingroup \@sanitize@url \@href}%
\providecommand \@href[1]{\@@startlink{#1}\@@href}%
\providecommand \@@href[1]{\endgroup#1\@@endlink}%
\providecommand \@sanitize@url [0]{\catcode `\\12\catcode `\$12\catcode
`\&12\catcode `\#12\catcode `\^12\catcode `\_12\catcode `\%12\relax}%
\providecommand \@@startlink[1]{}%
\providecommand \@@endlink[0]{}%
\providecommand \url [0]{\begingroup\@sanitize@url \@url }%
\providecommand \@url [1]{\endgroup\@href {#1}{\urlprefix }}%
\providecommand \urlprefix [0]{URL }%
\providecommand \Eprint [0]{\href }%
\providecommand \doibase [0]{http://dx.doi.org/}%
\providecommand \selectlanguage [0]{\@gobble}%
\providecommand \bibinfo [0]{\@secondoftwo}%
\providecommand \bibfield [0]{\@secondoftwo}%
\providecommand \translation [1]{[#1]}%
\providecommand \BibitemOpen [0]{}%
\providecommand \bibitemStop [0]{}%
\providecommand \bibitemNoStop [0]{.\EOS\space}%
\providecommand \EOS [0]{\spacefactor3000\relax}%
\providecommand \BibitemShut [1]{\csname bibitem#1\endcsname}%
\let\auto@bib@innerbib\@empty
\bibitem [{\citenamefont {Giunti}\ and\ \citenamefont
{Studenikin}(2015)}]{Giunti:2014ixa}%
\BibitemOpen
\bibfield {author} {\bibinfo {author} {\bibfnamefont {C.}~\bibnamefont
{Giunti}}\ and\ \bibinfo {author} {\bibfnamefont {A.}~\bibnamefont
{Studenikin}},\ }\href@noop {} {\bibfield {journal} {\bibinfo {journal}
{Rev.Mod.Phys.}\ }\textbf {\bibinfo {volume} {87}},\ \bibinfo {pages} {531}
(\bibinfo {year} {2015})},\ \Eprint {http://arxiv.org/abs/arXiv:1403.6344}
{arXiv:1403.6344 [hep-ph]} \BibitemShut {NoStop}%
\bibitem [{\citenamefont {Lee}(1972)}]{Lee:1973fw}%
\BibitemOpen
\bibfield {author} {\bibinfo {author} {\bibfnamefont {S.~Y.}\ \bibnamefont
{Lee}},\ }\href {\doibase 10.1103/PhysRevD.6.1701} {\bibfield {journal}
{\bibinfo {journal} {Phys. Rev.}\ }\textbf {\bibinfo {volume} {D6}},\
\bibinfo {pages} {1701} (\bibinfo {year} {1972})}\BibitemShut {NoStop}%
\bibitem [{\citenamefont {Lee}\ and\ \citenamefont
{Shrock}(1977)}]{Lee:1977tib}%
\BibitemOpen
\bibfield {author} {\bibinfo {author} {\bibfnamefont {B.~W.}\ \bibnamefont
{Lee}}\ and\ \bibinfo {author} {\bibfnamefont {R.~E.}\ \bibnamefont
{Shrock}},\ }\href {\doibase 10.1103/PhysRevD.16.1444} {\bibfield {journal}
{\bibinfo {journal} {Phys. Rev.}\ }\textbf {\bibinfo {volume} {D16}},\
\bibinfo {pages} {1444} (\bibinfo {year} {1977})}\BibitemShut {NoStop}%
\bibitem [{\citenamefont {Lucio}\ \emph {et~al.}(1984)\citenamefont {Lucio},
\citenamefont {Rosado},\ and\ \citenamefont {Zepeda}}]{Lucio:1983mg}%
\BibitemOpen
\bibfield {author} {\bibinfo {author} {\bibfnamefont {J.}~\bibnamefont
{Lucio}}, \bibinfo {author} {\bibfnamefont {A.}~\bibnamefont {Rosado}}, \
and\ \bibinfo {author} {\bibfnamefont {A.}~\bibnamefont {Zepeda}},\ }\href
{\doibase 10.1103/PhysRevD.29.1539} {\bibfield {journal} {\bibinfo
{journal} {Phys. Rev.}\ }\textbf {\bibinfo {volume} {D29}},\ \bibinfo {pages}
{1539} (\bibinfo {year} {1984})}\BibitemShut {NoStop}%
\bibitem [{\citenamefont {Lucio}\ \emph {et~al.}(1985)\citenamefont {Lucio},
\citenamefont {Rosado},\ and\ \citenamefont {Zepeda}}]{Lucio:1984jn}%
\BibitemOpen
\bibfield {author} {\bibinfo {author} {\bibfnamefont {J.}~\bibnamefont
{Lucio}}, \bibinfo {author} {\bibfnamefont {A.}~\bibnamefont {Rosado}}, \
and\ \bibinfo {author} {\bibfnamefont {A.}~\bibnamefont {Zepeda}},\ }\href
{\doibase 10.1103/PhysRevD.31.1091} {\bibfield {journal} {\bibinfo
{journal} {Phys. Rev.}\ }\textbf {\bibinfo {volume} {D31}},\ \bibinfo {pages}
{1091} (\bibinfo {year} {1985})}\BibitemShut {NoStop}%
\bibitem [{\citenamefont {Degrassi}\ \emph {et~al.}(1989)\citenamefont
{Degrassi}, \citenamefont {Sirlin},\ and\ \citenamefont
{Marciano}}]{Degrassi:1989ip}%
\BibitemOpen
\bibfield {author} {\bibinfo {author} {\bibfnamefont {G.}~\bibnamefont
{Degrassi}}, \bibinfo {author} {\bibfnamefont {A.}~\bibnamefont {Sirlin}}, \
and\ \bibinfo {author} {\bibfnamefont {W.~J.}\ \bibnamefont {Marciano}},\
}\href {\doibase 10.1103/PhysRevD.39.287} {\bibfield {journal} {\bibinfo
{journal} {Phys. Rev.}\ }\textbf {\bibinfo {volume} {D39}},\ \bibinfo {pages}
{287} (\bibinfo {year} {1989})}\BibitemShut {NoStop}%
\bibitem [{\citenamefont {Papavassiliou}(1990)}]{Papavassiliou:1989zd}%
\BibitemOpen
\bibfield {author} {\bibinfo {author} {\bibfnamefont {J.}~\bibnamefont
{Papavassiliou}},\ }\href {\doibase 10.1103/PhysRevD.41.3179} {\bibfield
{journal} {\bibinfo {journal} {Phys. Rev.}\ }\textbf {\bibinfo {volume}
{D41}},\ \bibinfo {pages} {3179} (\bibinfo {year} {1990})}\BibitemShut
{NoStop}%
\bibitem [{\citenamefont {Bernabeu}\ \emph {et~al.}(2000)\citenamefont
{Bernabeu}, \citenamefont {Cabral-Rosetti}, \citenamefont {Papavassiliou},\
and\ \citenamefont {Vidal}}]{Bernabeu:2000hf}%
\BibitemOpen
\bibfield {author} {\bibinfo {author} {\bibfnamefont {J.}~\bibnamefont
{Bernabeu}}, \bibinfo {author} {\bibfnamefont {L.~G.}\ \bibnamefont
{Cabral-Rosetti}}, \bibinfo {author} {\bibfnamefont {J.}~\bibnamefont
{Papavassiliou}}, \ and\ \bibinfo {author} {\bibfnamefont {J.}~\bibnamefont
{Vidal}},\ }\href {\doibase 10.1103/PhysRevD.62.113012} {\bibfield {journal}
{\bibinfo {journal} {Phys. Rev.}\ }\textbf {\bibinfo {volume} {D62}},\
\bibinfo {pages} {113012} (\bibinfo {year} {2000})},\ \Eprint
{http://arxiv.org/abs/hep-ph/0008114} {hep-ph/0008114} \BibitemShut {NoStop}%
\bibitem [{\citenamefont {Bernabeu}\ \emph {et~al.}(2002)\citenamefont
{Bernabeu}, \citenamefont {Papavassiliou},\ and\ \citenamefont
{Vidal}}]{Bernabeu:2002nw}%
\BibitemOpen
\bibfield {author} {\bibinfo {author} {\bibfnamefont {J.}~\bibnamefont
{Bernabeu}}, \bibinfo {author} {\bibfnamefont {J.}~\bibnamefont
{Papavassiliou}}, \ and\ \bibinfo {author} {\bibfnamefont {J.}~\bibnamefont
{Vidal}},\ }\href@noop {} {\bibfield {journal} {\bibinfo {journal} {Phys.
Rev. Lett.}\ }\textbf {\bibinfo {volume} {89}},\ \bibinfo {pages} {101802}
(\bibinfo {year} {2002})},\ \Eprint {http://arxiv.org/abs/hep-ph/0206015}
{hep-ph/0206015} \BibitemShut {NoStop}%
\bibitem [{\citenamefont {Bernabeu}\ \emph {et~al.}(2004)\citenamefont
{Bernabeu}, \citenamefont {Papavassiliou},\ and\ \citenamefont
{Vidal}}]{Bernabeu:2002pd}%
\BibitemOpen
\bibfield {author} {\bibinfo {author} {\bibfnamefont {J.}~\bibnamefont
{Bernabeu}}, \bibinfo {author} {\bibfnamefont {J.}~\bibnamefont
{Papavassiliou}}, \ and\ \bibinfo {author} {\bibfnamefont {J.}~\bibnamefont
{Vidal}},\ }\href@noop {} {\bibfield {journal} {\bibinfo {journal} {Nucl.
Phys.}\ }\textbf {\bibinfo {volume} {B680}},\ \bibinfo {pages} {450}
(\bibinfo {year} {2004})},\ \Eprint {http://arxiv.org/abs/hep-ph/0210055}
{hep-ph/0210055} \BibitemShut {NoStop}%
\bibitem [{\citenamefont {Fujikawa}\ and\ \citenamefont
{Shrock}(2004)}]{Fujikawa:2003ww}%
\BibitemOpen
\bibfield {author} {\bibinfo {author} {\bibfnamefont {K.}~\bibnamefont
{Fujikawa}}\ and\ \bibinfo {author} {\bibfnamefont {R.}~\bibnamefont
{Shrock}},\ }\href@noop {} {\bibfield {journal} {\bibinfo {journal} {Phys.
Rev.}\ }\textbf {\bibinfo {volume} {D69}},\ \bibinfo {pages} {013007}
(\bibinfo {year} {2004})},\ \Eprint {http://arxiv.org/abs/hep-ph/0309329}
{hep-ph/0309329} \BibitemShut {NoStop}%
\bibitem [{\citenamefont {Papavassiliou}\ \emph {et~al.}(2004)\citenamefont
{Papavassiliou}, \citenamefont {Bernabeu}, \citenamefont {Binosi},\ and\
\citenamefont {Vidal}}]{Papavassiliou:2003rx}%
\BibitemOpen
\bibfield {author} {\bibinfo {author} {\bibfnamefont {J.}~\bibnamefont
{Papavassiliou}}, \bibinfo {author} {\bibfnamefont {J.}~\bibnamefont
{Bernabeu}}, \bibinfo {author} {\bibfnamefont {D.}~\bibnamefont {Binosi}}, \
and\ \bibinfo {author} {\bibfnamefont {J.}~\bibnamefont {Vidal}},\
}\href@noop {} {\bibfield {journal} {\bibinfo {journal} {Eur. Phys. J.}\
}\textbf {\bibinfo {volume} {C33}},\ \bibinfo {pages} {S865} (\bibinfo {year}
{2004})},\ \Eprint {http://arxiv.org/abs/hep-ph/0310028} {hep-ph/0310028}
\BibitemShut {NoStop}%
\bibitem [{\citenamefont {Bernabeu}\ \emph {et~al.}(2005)\citenamefont
{Bernabeu}, \citenamefont {Binosi},\ and\ \citenamefont
{Papavassiliou}}]{Bernabeu:2004jr}%
\BibitemOpen
\bibfield {author} {\bibinfo {author} {\bibfnamefont {J.}~\bibnamefont
{Bernabeu}}, \bibinfo {author} {\bibfnamefont {D.}~\bibnamefont {Binosi}}, \
and\ \bibinfo {author} {\bibfnamefont {J.}~\bibnamefont {Papavassiliou}},\
}\href@noop {} {\bibfield {journal} {\bibinfo {journal} {Nucl. Phys.}\
}\textbf {\bibinfo {volume} {B716}},\ \bibinfo {pages} {352} (\bibinfo {year}
{2005})},\ \Eprint {http://arxiv.org/abs/hep-ph/0405288} {hep-ph/0405288}
\BibitemShut {NoStop}%
\bibitem [{\citenamefont {Sehgal}(1985)}]{Sehgal:1985iu}%
\BibitemOpen
\bibfield {author} {\bibinfo {author} {\bibfnamefont {L.~M.}\ \bibnamefont
{Sehgal}},\ }\href {\doibase 10.1016/0370-2693(85)90942-6} {\bibfield
{journal} {\bibinfo {journal} {Phys. Lett.}\ }\textbf {\bibinfo {volume}
{162B}},\ \bibinfo {pages} {370} (\bibinfo {year} {1985})}\BibitemShut
{NoStop}%
\bibitem [{\citenamefont {Papavassiliou}\ \emph {et~al.}(2006)\citenamefont
{Papavassiliou}, \citenamefont {Bernabeu},\ and\ \citenamefont
{Passera}}]{Papavassiliou:2005cs}%
\BibitemOpen
\bibfield {author} {\bibinfo {author} {\bibfnamefont {J.}~\bibnamefont
{Papavassiliou}}, \bibinfo {author} {\bibfnamefont {J.}~\bibnamefont
{Bernabeu}}, \ and\ \bibinfo {author} {\bibfnamefont {M.}~\bibnamefont
{Passera}},\ }\href@noop {} {\bibfield {journal} {\bibinfo {journal} {PoS}\
}\textbf {\bibinfo {volume} {HEP2005}},\ \bibinfo {pages} {192} (\bibinfo
{year} {2006})},\ \Eprint {http://arxiv.org/abs/hep-ph/0512029}
{hep-ph/0512029} \BibitemShut {NoStop}%
\bibitem [{\citenamefont {Kosmas}\ \emph {et~al.}(2015)\citenamefont {Kosmas},
\citenamefont {Miranda}, \citenamefont {Papoulias}, \citenamefont {Tortola},\
and\ \citenamefont {Valle}}]{Kosmas:2015vsa}%
\BibitemOpen
\bibfield {author} {\bibinfo {author} {\bibfnamefont {T.~S.}\ \bibnamefont
{Kosmas}}, \bibinfo {author} {\bibfnamefont {O.~G.}\ \bibnamefont {Miranda}},
\bibinfo {author} {\bibfnamefont {D.~K.}\ \bibnamefont {Papoulias}}, \bibinfo
{author} {\bibfnamefont {M.}~\bibnamefont {Tortola}}, \ and\ \bibinfo
{author} {\bibfnamefont {J.~W.~F.}\ \bibnamefont {Valle}},\ }\href@noop {}
{\bibfield {journal} {\bibinfo {journal} {Phys. Lett.}\ }\textbf {\bibinfo
{volume} {B750}},\ \bibinfo {pages} {459} (\bibinfo {year} {2015})},\ \Eprint
{http://arxiv.org/abs/arXiv:1506.08377} {arXiv:1506.08377 [hep-ph]}
\BibitemShut {NoStop}%
\bibitem [{\citenamefont {Papoulias}\ and\ \citenamefont
{Kosmas}(2018)}]{Kosmas:2017tsq}%
\BibitemOpen
\bibfield {author} {\bibinfo {author} {\bibfnamefont {D.~K.}\ \bibnamefont
{Papoulias}}\ and\ \bibinfo {author} {\bibfnamefont {T.~S.}\ \bibnamefont
{Kosmas}},\ }\href@noop {} {\bibfield {journal} {\bibinfo {journal}
{Phys.Rev.}\ }\textbf {\bibinfo {volume} {D97}},\ \bibinfo {pages} {033003}
(\bibinfo {year} {2018})},\ \Eprint {http://arxiv.org/abs/arXiv:1711.09773}
{arXiv:1711.09773 [hep-ph]} \BibitemShut {NoStop}%
\bibitem [{\citenamefont {Akimov}\ \emph {et~al.}(2017)\citenamefont {Akimov}
\emph {et~al.}}]{Akimov:2017ade}%
\BibitemOpen
\bibfield {author} {\bibinfo {author} {\bibfnamefont {D.}~\bibnamefont
{Akimov}} \emph {et~al.} (\bibinfo {collaboration} {COHERENT}),\ }\href
{\doibase 10.1126/science.aao0990} {\bibfield {journal} {\bibinfo {journal}
{Science}\ }\textbf {\bibinfo {volume} {357}},\ \bibinfo {pages} {1123}
(\bibinfo {year} {2017})},\ \Eprint {http://arxiv.org/abs/arXiv:1708.01294}
{arXiv:1708.01294 [nucl-ex]} \BibitemShut {NoStop}%
\bibitem [{\citenamefont {Akimov}\ \emph
{et~al.}(2018{\natexlab{a}})\citenamefont {Akimov} \emph
{et~al.}}]{Akimov:2018vzs}%
\BibitemOpen
\bibfield {author} {\bibinfo {author} {\bibfnamefont {D.}~\bibnamefont
{Akimov}} \emph {et~al.} (\bibinfo {collaboration} {COHERENT}),\ }\href@noop
{} {\ (\bibinfo {year} {2018}{\natexlab{a}})},\ \Eprint
{http://arxiv.org/abs/arXiv:1804.09459} {arXiv:1804.09459 [nucl-ex]}
\BibitemShut {NoStop}%
\bibitem [{\citenamefont {Freedman}(1974)}]{Freedman:1973yd}%
\BibitemOpen
\bibfield {author} {\bibinfo {author} {\bibfnamefont {D.~Z.}\ \bibnamefont
{Freedman}},\ }\href {\doibase 10.1103/PhysRevD.9.1389} {\bibfield {journal}
{\bibinfo {journal} {Phys. Rev.}\ }\textbf {\bibinfo {volume} {D9}},\
\bibinfo {pages} {1389} (\bibinfo {year} {1974})}\BibitemShut {NoStop}%
\bibitem [{\citenamefont {Freedman}\ \emph {et~al.}(1977)\citenamefont
{Freedman}, \citenamefont {Schramm},\ and\ \citenamefont
{Tubbs}}]{Freedman:1977xn}%
\BibitemOpen
\bibfield {author} {\bibinfo {author} {\bibfnamefont {D.~Z.}\ \bibnamefont
{Freedman}}, \bibinfo {author} {\bibfnamefont {D.~N.}\ \bibnamefont
{Schramm}}, \ and\ \bibinfo {author} {\bibfnamefont {D.~L.}\ \bibnamefont
{Tubbs}},\ }\href {\doibase 10.1146/annurev.ns.27.120177.001123} {\bibfield
{journal} {\bibinfo {journal} {Ann. Rev. Nucl. Part. Sci.}\ }\textbf
{\bibinfo {volume} {27}},\ \bibinfo {pages} {167} (\bibinfo {year}
{1977})}\BibitemShut {NoStop}%
\bibitem [{\citenamefont {Drukier}\ and\ \citenamefont
{Stodolsky}(1984)}]{Drukier:1983gj}%
\BibitemOpen
\bibfield {author} {\bibinfo {author} {\bibfnamefont {A.}~\bibnamefont
{Drukier}}\ and\ \bibinfo {author} {\bibfnamefont {L.}~\bibnamefont
{Stodolsky}},\ }\href {\doibase 10.1103/PhysRevD.30.2295} {\bibfield
{journal} {\bibinfo {journal} {Phys. Rev.}\ }\textbf {\bibinfo {volume}
{D30}},\ \bibinfo {pages} {2295} (\bibinfo {year} {1984})}\BibitemShut
{NoStop}%
\bibitem [{\citenamefont {A.Bednyakov}\ and\ \citenamefont
{V.Naumov}(2018)}]{Bednyakov:2018mjd}%
\BibitemOpen
\bibfield {author} {\bibinfo {author} {\bibfnamefont {V.}~\bibnamefont
{A.Bednyakov}}\ and\ \bibinfo {author} {\bibfnamefont {D.}~\bibnamefont
{V.Naumov}},\ }\href@noop {} {\bibfield {journal} {\bibinfo {journal}
{Phys.Rev.}\ }\textbf {\bibinfo {volume} {D98}},\ \bibinfo {pages} {053004}
(\bibinfo {year} {2018})},\ \Eprint {http://arxiv.org/abs/arXiv:1806.08768}
{arXiv:1806.08768 [hep-ph]} \BibitemShut {NoStop}%
\bibitem [{\citenamefont {Barranco}\ \emph {et~al.}(2005)\citenamefont
{Barranco}, \citenamefont {Miranda},\ and\ \citenamefont
{Rashba}}]{Barranco:2005yy}%
\BibitemOpen
\bibfield {author} {\bibinfo {author} {\bibfnamefont {J.}~\bibnamefont
{Barranco}}, \bibinfo {author} {\bibfnamefont {O.~G.}\ \bibnamefont
{Miranda}}, \ and\ \bibinfo {author} {\bibfnamefont {T.~I.}\ \bibnamefont
{Rashba}},\ }\href@noop {} {\bibfield {journal} {\bibinfo {journal} {JHEP}\
}\textbf {\bibinfo {volume} {0512}},\ \bibinfo {pages} {021} (\bibinfo {year}
{2005})},\ \Eprint {http://arxiv.org/abs/hep-ph/0508299} {hep-ph/0508299}
\BibitemShut {NoStop}%
\bibitem [{\citenamefont {Patton}\ \emph {et~al.}(2012)\citenamefont {Patton},
\citenamefont {Engel}, \citenamefont {McLaughlin},\ and\ \citenamefont
{Schunck}}]{Patton:2012jr}%
\BibitemOpen
\bibfield {author} {\bibinfo {author} {\bibfnamefont {K.}~\bibnamefont
{Patton}}, \bibinfo {author} {\bibfnamefont {J.}~\bibnamefont {Engel}},
\bibinfo {author} {\bibfnamefont {G.~C.}\ \bibnamefont {McLaughlin}}, \ and\
\bibinfo {author} {\bibfnamefont {N.}~\bibnamefont {Schunck}},\ }\href
{\doibase 10.1103/PhysRevC.86.024612} {\bibfield {journal} {\bibinfo
{journal} {Phys. Rev.}\ }\textbf {\bibinfo {volume} {C86}},\ \bibinfo {pages}
{024612} (\bibinfo {year} {2012})},\ \Eprint
{http://arxiv.org/abs/arXiv:1207.0693} {arXiv:1207.0693 [nucl-th]}
\BibitemShut {NoStop}%
\bibitem [{\citenamefont {Papoulias}\ and\ \citenamefont
{Kosmas}(2015)}]{Papoulias:2015vxa}%
\BibitemOpen
\bibfield {author} {\bibinfo {author} {\bibfnamefont {D.~K.}\ \bibnamefont
{Papoulias}}\ and\ \bibinfo {author} {\bibfnamefont {T.~S.}\ \bibnamefont
{Kosmas}},\ }\href {\doibase 10.1155/2015/763648} {\bibfield {journal}
{\bibinfo {journal} {Adv. High Energy Phys.}\ }\textbf {\bibinfo {volume}
{2015}},\ \bibinfo {pages} {763648} (\bibinfo {year} {2015})},\ \Eprint
{http://arxiv.org/abs/arXiv:1502.02928} {arXiv:1502.02928 [nucl-th]}
\BibitemShut {NoStop}%
\bibitem [{\citenamefont {Lindner}\ \emph {et~al.}(2017)\citenamefont
{Lindner}, \citenamefont {Rodejohann},\ and\ \citenamefont
{Xu}}]{Lindner:2016wff}%
\BibitemOpen
\bibfield {author} {\bibinfo {author} {\bibfnamefont {M.}~\bibnamefont
{Lindner}}, \bibinfo {author} {\bibfnamefont {W.}~\bibnamefont {Rodejohann}},
\ and\ \bibinfo {author} {\bibfnamefont {X.-J.}\ \bibnamefont {Xu}},\
}\href@noop {} {\bibfield {journal} {\bibinfo {journal} {JHEP}\ }\textbf
{\bibinfo {volume} {1703}},\ \bibinfo {pages} {097} (\bibinfo {year}
{2017})},\ \Eprint {http://arxiv.org/abs/arXiv:1612.04150} {arXiv:1612.04150
[hep-ph]} \BibitemShut {NoStop}%
\bibitem [{\citenamefont {Shoemaker}(2017)}]{Shoemaker:2017lzs}%
\BibitemOpen
\bibfield {author} {\bibinfo {author} {\bibfnamefont {I.~M.}\ \bibnamefont
{Shoemaker}},\ }\href@noop {} {\bibfield {journal} {\bibinfo {journal}
{Phys.Rev.}\ }\textbf {\bibinfo {volume} {D95}},\ \bibinfo {pages} {115028}
(\bibinfo {year} {2017})},\ \Eprint {http://arxiv.org/abs/arXiv:1703.05774}
{arXiv:1703.05774 [hep-ph]} \BibitemShut {NoStop}%
\bibitem [{\citenamefont {Ciuffoli}\ \emph {et~al.}(2018)\citenamefont
{Ciuffoli}, \citenamefont {Evslin}, \citenamefont {Fu},\ and\ \citenamefont
{Tang}}]{Ciuffoli:2018qem}%
\BibitemOpen
\bibfield {author} {\bibinfo {author} {\bibfnamefont {E.}~\bibnamefont
{Ciuffoli}}, \bibinfo {author} {\bibfnamefont {J.}~\bibnamefont {Evslin}},
\bibinfo {author} {\bibfnamefont {Q.}~\bibnamefont {Fu}}, \ and\ \bibinfo
{author} {\bibfnamefont {J.}~\bibnamefont {Tang}},\ }\href@noop {} {\bibfield
{journal} {\bibinfo {journal} {Phys.Rev.}\ }\textbf {\bibinfo {volume}
{D97}},\ \bibinfo {pages} {113003} (\bibinfo {year} {2018})},\ \Eprint
{http://arxiv.org/abs/arXiv:1801.02166} {arXiv:1801.02166 [physics]}
\BibitemShut {NoStop}%
\bibitem [{\citenamefont {Canas}\ \emph {et~al.}(2018)\citenamefont {Canas},
\citenamefont {Garces}, \citenamefont {Miranda},\ and\ \citenamefont
{Parada}}]{Canas:2018rng}%
\BibitemOpen
\bibfield {author} {\bibinfo {author} {\bibfnamefont {B.~C.}\ \bibnamefont
{Canas}}, \bibinfo {author} {\bibfnamefont {E.~A.}\ \bibnamefont {Garces}},
\bibinfo {author} {\bibfnamefont {O.~G.}\ \bibnamefont {Miranda}}, \ and\
\bibinfo {author} {\bibfnamefont {A.}~\bibnamefont {Parada}},\ }\href@noop {}
{\bibfield {journal} {\bibinfo {journal} {Phys.Lett.}\ }\textbf {\bibinfo
{volume} {B784}},\ \bibinfo {pages} {159} (\bibinfo {year} {2018})},\ \Eprint
{http://arxiv.org/abs/arXiv:1806.01310} {arXiv:1806.01310 [hep-ph]}
\BibitemShut {NoStop}%
\bibitem [{\citenamefont {Billard}\ \emph {et~al.}(2018)\citenamefont
{Billard}, \citenamefont {Johnston},\ and\ \citenamefont
{Kavanagh}}]{Billard:2018jnl}%
\BibitemOpen
\bibfield {author} {\bibinfo {author} {\bibfnamefont {J.}~\bibnamefont
{Billard}}, \bibinfo {author} {\bibfnamefont {J.}~\bibnamefont {Johnston}}, \
and\ \bibinfo {author} {\bibfnamefont {B.~J.}\ \bibnamefont {Kavanagh}},\
}\href@noop {} {\bibfield {journal} {\bibinfo {journal} {JCAP}\ }\textbf
{\bibinfo {volume} {1811}},\ \bibinfo {pages} {016} (\bibinfo {year}
{2018})},\ \Eprint {http://arxiv.org/abs/arXiv:1805.01798} {arXiv:1805.01798
[hep-ph]} \BibitemShut {NoStop}%
\bibitem [{\citenamefont {Brdar}\ \emph {et~al.}(2018)\citenamefont {Brdar},
\citenamefont {Rodejohann},\ and\ \citenamefont {Xu}}]{Brdar:2018qqj}%
\BibitemOpen
\bibfield {author} {\bibinfo {author} {\bibfnamefont {V.}~\bibnamefont
{Brdar}}, \bibinfo {author} {\bibfnamefont {W.}~\bibnamefont {Rodejohann}}, \
and\ \bibinfo {author} {\bibfnamefont {X.-J.}\ \bibnamefont {Xu}},\
}\href@noop {} {\bibfield {journal} {\bibinfo {journal} {JHEP}\ }\textbf
{\bibinfo {volume} {1812}},\ \bibinfo {pages} {024} (\bibinfo {year}
{2018})},\ \Eprint {http://arxiv.org/abs/arXiv:1810.03626} {arXiv:1810.03626
[hep-ph]} \BibitemShut {NoStop}%
\bibitem [{\citenamefont {Cadeddu}\ \emph {et~al.}(2018)\citenamefont
{Cadeddu}, \citenamefont {Giunti}, \citenamefont {Li},\ and\ \citenamefont
{Zhang}}]{Cadeddu:2017etk}%
\BibitemOpen
\bibfield {author} {\bibinfo {author} {\bibfnamefont {M.}~\bibnamefont
{Cadeddu}}, \bibinfo {author} {\bibfnamefont {C.}~\bibnamefont {Giunti}},
\bibinfo {author} {\bibfnamefont {Y.~F.}\ \bibnamefont {Li}}, \ and\ \bibinfo
{author} {\bibfnamefont {Y.~Y.}\ \bibnamefont {Zhang}},\ }\href@noop {}
{\bibfield {journal} {\bibinfo {journal} {Phys.Rev.Lett.}\ }\textbf
{\bibinfo {volume} {120}},\ \bibinfo {pages} {072501} (\bibinfo {year}
{2018})},\ \Eprint {http://arxiv.org/abs/arXiv:1710.02730} {arXiv:1710.02730
[hep-ph]} \BibitemShut {NoStop}%
\bibitem [{\citenamefont {Coloma}\ \emph {et~al.}(2017)\citenamefont {Coloma},
\citenamefont {Gonzalez-Garcia}, \citenamefont {Maltoni},\ and\ \citenamefont
{Schwetz}}]{Coloma:2017ncl}%
\BibitemOpen
\bibfield {author} {\bibinfo {author} {\bibfnamefont {P.}~\bibnamefont
{Coloma}}, \bibinfo {author} {\bibfnamefont {M.~C.}\ \bibnamefont
{Gonzalez-Garcia}}, \bibinfo {author} {\bibfnamefont {M.}~\bibnamefont
{Maltoni}}, \ and\ \bibinfo {author} {\bibfnamefont {T.}~\bibnamefont
{Schwetz}},\ }\href@noop {} {\bibfield {journal} {\bibinfo {journal}
{Phys.Rev.}\ }\textbf {\bibinfo {volume} {D96}},\ \bibinfo {pages} {115007}
(\bibinfo {year} {2017})},\ \Eprint {http://arxiv.org/abs/arXiv:1708.02899}
{arXiv:1708.02899 [hep-ph]} \BibitemShut {NoStop}%
\bibitem [{\citenamefont {Liao}\ and\ \citenamefont
{Marfatia}(2017)}]{Liao:2017uzy}%
\BibitemOpen
\bibfield {author} {\bibinfo {author} {\bibfnamefont {J.}~\bibnamefont
{Liao}}\ and\ \bibinfo {author} {\bibfnamefont {D.}~\bibnamefont
{Marfatia}},\ }\href@noop {} {\bibfield {journal} {\bibinfo {journal}
{Phys.Lett.}\ }\textbf {\bibinfo {volume} {B775}},\ \bibinfo {pages} {54}
(\bibinfo {year} {2017})},\ \Eprint {http://arxiv.org/abs/arXiv:1708.04255}
{arXiv:1708.04255 [hep-ph]} \BibitemShut {NoStop}%
\bibitem [{\citenamefont {Denton}\ \emph {et~al.}(2018)\citenamefont {Denton},
\citenamefont {Farzan},\ and\ \citenamefont {Shoemaker}}]{Denton:2018xmq}%
\BibitemOpen
\bibfield {author} {\bibinfo {author} {\bibfnamefont {P.~B.}\ \bibnamefont
{Denton}}, \bibinfo {author} {\bibfnamefont {Y.}~\bibnamefont {Farzan}}, \
and\ \bibinfo {author} {\bibfnamefont {I.~M.}\ \bibnamefont {Shoemaker}},\
}\href@noop {} {\bibfield {journal} {\bibinfo {journal} {JHEP}\ }\textbf
{\bibinfo {volume} {1807}},\ \bibinfo {pages} {037} (\bibinfo {year}
{2018})},\ \Eprint {http://arxiv.org/abs/arXiv:1804.03660} {arXiv:1804.03660
[hep-ph]} \BibitemShut {NoStop}%
\bibitem [{\citenamefont {Aristizabal~Sierra}\ \emph
{et~al.}(2018)\citenamefont {Aristizabal~Sierra}, \citenamefont {De~Romeri},\
and\ \citenamefont {Rojas}}]{AristizabalSierra:2018eqm}%
\BibitemOpen
\bibfield {author} {\bibinfo {author} {\bibfnamefont {D.}~\bibnamefont
{Aristizabal~Sierra}}, \bibinfo {author} {\bibfnamefont {V.}~\bibnamefont
{De~Romeri}}, \ and\ \bibinfo {author} {\bibfnamefont {N.}~\bibnamefont
{Rojas}},\ }\href@noop {} {\bibfield {journal} {\bibinfo {journal}
{Phys.Rev.}\ }\textbf {\bibinfo {volume} {D98}},\ \bibinfo {pages} {075018}
(\bibinfo {year} {2018})},\ \Eprint {http://arxiv.org/abs/arXiv:1806.07424}
{arXiv:1806.07424 [hep-ph]} \BibitemShut {NoStop}%
\bibitem [{\citenamefont {Cadeddu}\ and\ \citenamefont
{Dordei}(2019)}]{Cadeddu:2018izq}%
\BibitemOpen
\bibfield {author} {\bibinfo {author} {\bibfnamefont {M.}~\bibnamefont
{Cadeddu}}\ and\ \bibinfo {author} {\bibfnamefont {F.}~\bibnamefont
{Dordei}},\ }\href@noop {} {\bibfield {journal} {\bibinfo {journal}
{Phys.Rev.}\ }\textbf {\bibinfo {volume} {D99}},\ \bibinfo {pages} {033010}
(\bibinfo {year} {2019})},\ \Eprint {http://arxiv.org/abs/arXiv:1808.10202}
{arXiv:1808.10202 [hep-ph]} \BibitemShut {NoStop}%
\bibitem [{\citenamefont {Bardeen}\ \emph {et~al.}(1972)\citenamefont
{Bardeen}, \citenamefont {Gastmans},\ and\ \citenamefont
{Lautrup}}]{Bardeen:1972vi}%
\BibitemOpen
\bibfield {author} {\bibinfo {author} {\bibfnamefont {W.~A.}\ \bibnamefont
{Bardeen}}, \bibinfo {author} {\bibfnamefont {R.}~\bibnamefont {Gastmans}}, \
and\ \bibinfo {author} {\bibfnamefont {B.~E.}\ \bibnamefont {Lautrup}},\
}\href@noop {} {\bibfield {journal} {\bibinfo {journal} {Nucl. Phys.}\
}\textbf {\bibinfo {volume} {B46}},\ \bibinfo {pages} {319} (\bibinfo {year}
{1972})}\BibitemShut {NoStop}%
\bibitem [{\citenamefont {Dvornikov}\ and\ \citenamefont
{Studenikin}(2004{\natexlab{a}})}]{Dvornikov:2003js}%
\BibitemOpen
\bibfield {author} {\bibinfo {author} {\bibfnamefont {M.}~\bibnamefont
{Dvornikov}}\ and\ \bibinfo {author} {\bibfnamefont {A.}~\bibnamefont
{Studenikin}},\ }\href@noop {} {\bibfield {journal} {\bibinfo {journal}
{Phys. Rev.}\ }\textbf {\bibinfo {volume} {D69}},\ \bibinfo {pages} {073001}
(\bibinfo {year} {2004}{\natexlab{a}})},\ \Eprint
{http://arxiv.org/abs/hep-ph/0305206} {hep-ph/0305206} \BibitemShut {NoStop}%
\bibitem [{\citenamefont {Dvornikov}\ and\ \citenamefont
{Studenikin}(2004{\natexlab{b}})}]{Dvornikov:2004sj}%
\BibitemOpen
\bibfield {author} {\bibinfo {author} {\bibfnamefont {M.}~\bibnamefont
{Dvornikov}}\ and\ \bibinfo {author} {\bibfnamefont {A.}~\bibnamefont
{Studenikin}},\ }\href@noop {} {\bibfield {journal} {\bibinfo {journal} {J.
Exp. Theor. Phys.}\ }\textbf {\bibinfo {volume} {99}},\ \bibinfo {pages}
{254} (\bibinfo {year} {2004}{\natexlab{b}})},\ \Eprint
{http://arxiv.org/abs/hep-ph/0411085} {hep-ph/0411085} \BibitemShut {NoStop}%
\bibitem [{\citenamefont {Tanabashi}\ \emph {et~al.}(2018)\citenamefont
{Tanabashi} \emph {et~al.}}]{Tanabashi:2018oca}%
\BibitemOpen
\bibfield {author} {\bibinfo {author} {\bibfnamefont {M.}~\bibnamefont
{Tanabashi}} \emph {et~al.} (\bibinfo {collaboration} {Particle Data
Group}),\ }\href {\doibase 10.1103/PhysRevD.98.030001} {\bibfield {journal}
{\bibinfo {journal} {Phys. Rev.}\ }\textbf {\bibinfo {volume} {D98}},\
\bibinfo {pages} {030001} (\bibinfo {year} {2018})}\BibitemShut {NoStop}%
\bibitem [{\citenamefont {Hirsch}\ \emph {et~al.}(2003)\citenamefont {Hirsch},
\citenamefont {Nardi},\ and\ \citenamefont {Restrepo}}]{Hirsch:2002uv}%
\BibitemOpen
\bibfield {author} {\bibinfo {author} {\bibfnamefont {M.}~\bibnamefont
{Hirsch}}, \bibinfo {author} {\bibfnamefont {E.}~\bibnamefont {Nardi}}, \
and\ \bibinfo {author} {\bibfnamefont {D.}~\bibnamefont {Restrepo}},\ }\href
{\doibase 10.1103/PhysRevD.67.033005} {\bibfield {journal} {\bibinfo
{journal} {Phys. Rev.}\ }\textbf {\bibinfo {volume} {D67}},\ \bibinfo {pages}
{033005} (\bibinfo {year} {2003})},\ \Eprint
{http://arxiv.org/abs/hep-ph/0210137} {hep-ph/0210137 [hep-ph]} \BibitemShut
{NoStop}%
\bibitem [{\citenamefont {Vidyakin}\ \emph {et~al.}(1992)\citenamefont
{Vidyakin}, \citenamefont {Vyrodov}, \citenamefont {Gurevich}, \citenamefont
{Kozlov}, \citenamefont {Martemyanov}, \citenamefont {Sukhotin},
\citenamefont {Tarasenkov}, \citenamefont {Turbin},\ and\ \citenamefont
{Khakhimov}}]{Vidyakin:1992nf}%
\BibitemOpen
\bibfield {author} {\bibinfo {author} {\bibfnamefont {G.~S.}\ \bibnamefont
{Vidyakin}}, \bibinfo {author} {\bibfnamefont {V.~N.}\ \bibnamefont
{Vyrodov}}, \bibinfo {author} {\bibfnamefont {I.~I.}\ \bibnamefont
{Gurevich}}, \bibinfo {author} {\bibfnamefont {Y.~V.}\ \bibnamefont
{Kozlov}}, \bibinfo {author} {\bibfnamefont {V.~P.}\ \bibnamefont
{Martemyanov}}, \bibinfo {author} {\bibfnamefont {S.~V.}\ \bibnamefont
{Sukhotin}}, \bibinfo {author} {\bibfnamefont {V.~G.}\ \bibnamefont
{Tarasenkov}}, \bibinfo {author} {\bibfnamefont {E.~V.}\ \bibnamefont
{Turbin}}, \ and\ \bibinfo {author} {\bibfnamefont {S.~K.}\ \bibnamefont
{Khakhimov}},\ }\href@noop {} {\bibfield {journal} {\bibinfo {journal}
{JETP Lett.}\ }\textbf {\bibinfo {volume} {55}},\ \bibinfo {pages} {206}
(\bibinfo {year} {1992})}\BibitemShut {NoStop}%
\bibitem [{\citenamefont {Deniz}\ \emph {et~al.}(2010)\citenamefont {Deniz}
\emph {et~al.}}]{Deniz:2009mu}%
\BibitemOpen
\bibfield {author} {\bibinfo {author} {\bibfnamefont {M.}~\bibnamefont
{Deniz}} \emph {et~al.} (\bibinfo {collaboration} {TEXONO}),\ }\href@noop {}
{\bibfield {journal} {\bibinfo {journal} {Phys. Rev.}\ }\textbf {\bibinfo
{volume} {D81}},\ \bibinfo {pages} {072001} (\bibinfo {year} {2010})},\
\Eprint {http://arxiv.org/abs/arXiv:0911.1597} {arXiv:0911.1597 [hep-ex]}
\BibitemShut {NoStop}%
\bibitem [{\citenamefont {Allen}\ \emph {et~al.}(1993)\citenamefont {Allen},
\citenamefont {Chen}, \citenamefont {Doe}, \citenamefont {Hausammann},
\citenamefont {Lee} \emph {et~al.}}]{Allen:1992qe}%
\BibitemOpen
\bibfield {author} {\bibinfo {author} {\bibfnamefont {R.}~\bibnamefont
{Allen}}, \bibinfo {author} {\bibfnamefont {H.}~\bibnamefont {Chen}},
\bibinfo {author} {\bibfnamefont {P.}~\bibnamefont {Doe}}, \bibinfo {author}
{\bibfnamefont {R.}~\bibnamefont {Hausammann}}, \bibinfo {author}
{\bibfnamefont {W.}~\bibnamefont {Lee}}, \emph {et~al.},\ }\href {\doibase
10.1103/PhysRevD.47.11} {\bibfield {journal} {\bibinfo {journal} {Phys.
Rev.}\ }\textbf {\bibinfo {volume} {D47}},\ \bibinfo {pages} {11} (\bibinfo
{year} {1993})}\BibitemShut {NoStop}%
\bibitem [{\citenamefont {Auerbach}\ \emph {et~al.}(2001)\citenamefont
{Auerbach} \emph {et~al.}}]{Auerbach:2001wg}%
\BibitemOpen
\bibfield {author} {\bibinfo {author} {\bibfnamefont {L.~B.}\ \bibnamefont
{Auerbach}} \emph {et~al.} (\bibinfo {collaboration} {LSND}),\ }\href@noop {}
{\bibfield {journal} {\bibinfo {journal} {Phys. Rev.}\ }\textbf {\bibinfo
{volume} {D63}},\ \bibinfo {pages} {112001} (\bibinfo {year} {2001})},\
\Eprint {http://arxiv.org/abs/hep-ex/0101039} {hep-ex/0101039} \BibitemShut
{NoStop}%
\bibitem [{\citenamefont {Ahrens}\ \emph {et~al.}(1990)\citenamefont {Ahrens},
\citenamefont {Aronson}, \citenamefont {Connolly}, \citenamefont {Gibbard},
\citenamefont {Murtagh} \emph {et~al.}}]{Ahrens:1990fp}%
\BibitemOpen
\bibfield {author} {\bibinfo {author} {\bibfnamefont {L.}~\bibnamefont
{Ahrens}}, \bibinfo {author} {\bibfnamefont {S.}~\bibnamefont {Aronson}},
\bibinfo {author} {\bibfnamefont {P.}~\bibnamefont {Connolly}}, \bibinfo
{author} {\bibfnamefont {B.}~\bibnamefont {Gibbard}}, \bibinfo {author}
{\bibfnamefont {M.}~\bibnamefont {Murtagh}}, \emph {et~al.},\ }\href
{\doibase 10.1103/PhysRevD.41.3297} {\bibfield {journal} {\bibinfo
{journal} {Phys. Rev.}\ }\textbf {\bibinfo {volume} {D41}},\ \bibinfo {pages}
{3297} (\bibinfo {year} {1990})}\BibitemShut {NoStop}%
\bibitem [{\citenamefont {Vilain}\ \emph {et~al.}(1995)\citenamefont {Vilain}
\emph {et~al.}}]{Vilain:1994hm}%
\BibitemOpen
\bibfield {author} {\bibinfo {author} {\bibfnamefont {P.}~\bibnamefont
{Vilain}} \emph {et~al.} (\bibinfo {collaboration} {CHARM-II}),\ }\href@noop
{} {\bibfield {journal} {\bibinfo {journal} {Phys. Lett.}\ }\textbf
{\bibinfo {volume} {B345}},\ \bibinfo {pages} {115} (\bibinfo {year}
{1995})}\BibitemShut {NoStop}%
\bibitem [{\citenamefont {Vogel}\ and\ \citenamefont
{Engel}(1989)}]{Vogel:1989iv}%
\BibitemOpen
\bibfield {author} {\bibinfo {author} {\bibfnamefont {P.}~\bibnamefont
{Vogel}}\ and\ \bibinfo {author} {\bibfnamefont {J.}~\bibnamefont {Engel}},\
}\href@noop {} {\bibfield {journal} {\bibinfo {journal} {Phys. Rev.}\
}\textbf {\bibinfo {volume} {D39}},\ \bibinfo {pages} {3378} (\bibinfo {year}
{1989})}\BibitemShut {NoStop}%
\bibitem [{\citenamefont {Kouzakov}\ and\ \citenamefont
{Studenikin}(2017)}]{Kouzakov:2017hbc}%
\BibitemOpen
\bibfield {author} {\bibinfo {author} {\bibfnamefont {K.~A.}\ \bibnamefont
{Kouzakov}}\ and\ \bibinfo {author} {\bibfnamefont {A.~I.}\ \bibnamefont
{Studenikin}},\ }\href@noop {} {\bibfield {journal} {\bibinfo {journal}
{Phys.Rev.}\ }\textbf {\bibinfo {volume} {D95}},\ \bibinfo {pages} {055013}
(\bibinfo {year} {2017})},\ \Eprint {http://arxiv.org/abs/arXiv:1703.00401}
{arXiv:1703.00401 [hep-ph]} \BibitemShut {NoStop}%
\bibitem [{\citenamefont {Barranco}\ \emph {et~al.}(2008)\citenamefont
{Barranco}, \citenamefont {Miranda},\ and\ \citenamefont
{Rashba}}]{Barranco:2007ea}%
\BibitemOpen
\bibfield {author} {\bibinfo {author} {\bibfnamefont {J.}~\bibnamefont
{Barranco}}, \bibinfo {author} {\bibfnamefont {O.~G.}\ \bibnamefont
{Miranda}}, \ and\ \bibinfo {author} {\bibfnamefont {T.~I.}\ \bibnamefont
{Rashba}},\ }\href@noop {} {\bibfield {journal} {\bibinfo {journal} {Phys.
Lett.}\ }\textbf {\bibinfo {volume} {B662}},\ \bibinfo {pages} {431}
(\bibinfo {year} {2008})},\ \Eprint {http://arxiv.org/abs/arXiv:0707.4319}
{arXiv:0707.4319 [hep-ph]} \BibitemShut {NoStop}%
\bibitem [{\citenamefont {Grau}\ and\ \citenamefont
{Grifols}(1986)}]{Grau:1985cn}%
\BibitemOpen
\bibfield {author} {\bibinfo {author} {\bibfnamefont {A.}~\bibnamefont
{Grau}}\ and\ \bibinfo {author} {\bibfnamefont {J.}~\bibnamefont {Grifols}},\
}\href {\doibase 10.1016/0370-2693(86)91385-7} {\bibfield {journal}
{\bibinfo {journal} {Phys.Lett.}\ }\textbf {\bibinfo {volume} {B166}},\
\bibinfo {pages} {233} (\bibinfo {year} {1986})}\BibitemShut {NoStop}%
\bibitem [{\citenamefont {Sakakibara}\ and\ \citenamefont
{Sehgal}(1979)}]{Sakakibara:1979rc}%
\BibitemOpen
\bibfield {author} {\bibinfo {author} {\bibfnamefont {S.}~\bibnamefont
{Sakakibara}}\ and\ \bibinfo {author} {\bibfnamefont {L.~M.}\ \bibnamefont
{Sehgal}},\ }\href {\doibase 10.1016/0370-2693(79)90893-1} {\bibfield
{journal} {\bibinfo {journal} {Phys. Lett.}\ }\textbf {\bibinfo {volume}
{83B}},\ \bibinfo {pages} {77} (\bibinfo {year} {1979})}\BibitemShut
{NoStop}%
\bibitem [{\citenamefont {Erler}\ and\ \citenamefont
{Su}(2013)}]{Erler:2013xha}%
\BibitemOpen
\bibfield {author} {\bibinfo {author} {\bibfnamefont {J.}~\bibnamefont
{Erler}}\ and\ \bibinfo {author} {\bibfnamefont {S.}~\bibnamefont {Su}},\
}\href {\doibase 10.1016/j.ppnp.2013.03.004} {\bibfield {journal} {\bibinfo
{journal} {Prog. Part. Nucl. Phys.}\ }\textbf {\bibinfo {volume} {71}},\
\bibinfo {pages} {119} (\bibinfo {year} {2013})},\ \Eprint
{http://arxiv.org/abs/arXiv:1303.5522} {arXiv:1303.5522 [hep-ph]}
\BibitemShut {NoStop}%
\bibitem [{\citenamefont {Zeller}\ \emph {et~al.}(2002)\citenamefont {Zeller}
\emph {et~al.}}]{Zeller:2001hh}%
\BibitemOpen
\bibfield {author} {\bibinfo {author} {\bibfnamefont {G.~P.}\ \bibnamefont
{Zeller}} \emph {et~al.} (\bibinfo {collaboration} {NuTeV}),\ }\href@noop {}
{\bibfield {journal} {\bibinfo {journal} {Phys. Rev. Lett.}\ }\textbf
{\bibinfo {volume} {88}},\ \bibinfo {pages} {091802} (\bibinfo {year}
{2002})},\ \Eprint {http://arxiv.org/abs/hep-ex/0110059} {hep-ex/0110059}
\BibitemShut {NoStop}%
\bibitem [{\citenamefont {Bahcall}\ \emph {et~al.}(1995)\citenamefont
{Bahcall}, \citenamefont {Kamionkowski},\ and\ \citenamefont
{Sirlin}}]{Bahcall:1995mm}%
\BibitemOpen
\bibfield {author} {\bibinfo {author} {\bibfnamefont {J.~N.}\ \bibnamefont
{Bahcall}}, \bibinfo {author} {\bibfnamefont {M.}~\bibnamefont
{Kamionkowski}}, \ and\ \bibinfo {author} {\bibfnamefont {A.}~\bibnamefont
{Sirlin}},\ }\href@noop {} {\bibfield {journal} {\bibinfo {journal} {Phys.
Rev.}\ }\textbf {\bibinfo {volume} {D51}},\ \bibinfo {pages} {6146} (\bibinfo
{year} {1995})},\ \Eprint {http://arxiv.org/abs/astro-ph/9502003}
{astro-ph/9502003} \BibitemShut {NoStop}%
\bibitem [{\citenamefont {Schael}\ \emph {et~al.}(2006)\citenamefont {Schael}
\emph {et~al.}}]{ALEPH:2005ab}%
\BibitemOpen
\bibfield {author} {\bibinfo {author} {\bibfnamefont {S.}~\bibnamefont
{Schael}} \emph {et~al.} (\bibinfo {collaboration} {ALEPH, DELPHI, L3, OPAL,
SLD, LEP Electroweak Working Group, SLD Electroweak Group, SLD Heavy Flavour
Group}),\ }\href@noop {} {\bibfield {journal} {\bibinfo {journal} {Phys.
Rept.}\ }\textbf {\bibinfo {volume} {427}},\ \bibinfo {pages} {257} (\bibinfo
{year} {2006})},\ \Eprint {http://arxiv.org/abs/hep-ex/0509008}
{hep-ex/0509008} \BibitemShut {NoStop}%
\bibitem [{\citenamefont {McFarland}\ \emph {et~al.}(1998)\citenamefont
{McFarland} \emph {et~al.}}]{McFarland:1997wx}%
\BibitemOpen
\bibfield {author} {\bibinfo {author} {\bibfnamefont {K.~S.}\ \bibnamefont
{McFarland}} \emph {et~al.} (\bibinfo {collaboration} {CCFR}),\ }\href@noop
{} {\bibfield {journal} {\bibinfo {journal} {Eur. Phys. J.}\ }\textbf
{\bibinfo {volume} {C1}},\ \bibinfo {pages} {509} (\bibinfo {year} {1998})},\
\Eprint {http://arxiv.org/abs/hep-ex/9701010} {hep-ex/9701010} \BibitemShut
{NoStop}%
\bibitem [{\citenamefont {Piekarewicz}\ \emph {et~al.}(2016)\citenamefont
{Piekarewicz}, \citenamefont {Linero}, \citenamefont {Giuliani},\ and\
\citenamefont {Chicken}}]{Piekarewicz:2016vbn}%
\BibitemOpen
\bibfield {author} {\bibinfo {author} {\bibfnamefont {J.}~\bibnamefont
{Piekarewicz}}, \bibinfo {author} {\bibfnamefont {A.~R.}\ \bibnamefont
{Linero}}, \bibinfo {author} {\bibfnamefont {P.}~\bibnamefont {Giuliani}}, \
and\ \bibinfo {author} {\bibfnamefont {E.}~\bibnamefont {Chicken}},\ }\href
{\doibase 10.1103/PhysRevC.94.034316} {\bibfield {journal} {\bibinfo
{journal} {Phys. Rev.}\ }\textbf {\bibinfo {volume} {C94}},\ \bibinfo {pages}
{034316} (\bibinfo {year} {2016})},\ \Eprint
{http://arxiv.org/abs/arXiv:1604.07799} {arXiv:1604.07799 [nucl-th]}
\BibitemShut {NoStop}%
\bibitem [{\citenamefont {Helm}(1956)}]{Helm:1956zz}%
\BibitemOpen
\bibfield {author} {\bibinfo {author} {\bibfnamefont {R.~H.}\ \bibnamefont
{Helm}},\ }\href {\doibase 10.1103/PhysRev.104.1466} {\bibfield {journal}
{\bibinfo {journal} {Phys. Rev.}\ }\textbf {\bibinfo {volume} {104}},\
\bibinfo {pages} {1466} (\bibinfo {year} {1956})}\BibitemShut {NoStop}%
\bibitem [{\citenamefont {Friedrich}\ and\ \citenamefont
{Voegler}(1982)}]{Friedrich:1982esq}%
\BibitemOpen
\bibfield {author} {\bibinfo {author} {\bibfnamefont {J.}~\bibnamefont
{Friedrich}}\ and\ \bibinfo {author} {\bibfnamefont {N.}~\bibnamefont
{Voegler}},\ }\href {\doibase 10.1016/0375-9474(82)90147-6} {\bibfield
{journal} {\bibinfo {journal} {Nucl. Phys.}\ }\textbf {\bibinfo {volume}
{A373}},\ \bibinfo {pages} {192} (\bibinfo {year} {1982})}\BibitemShut
{NoStop}%
\bibitem [{\citenamefont {Fricke}\ \emph {et~al.}(1995)\citenamefont {Fricke},
\citenamefont {Bernhardt}, \citenamefont {Heilig}, \citenamefont {Schaller},
\citenamefont {Schellenberg}, \citenamefont {Shera},\ and\ \citenamefont
{de~Jager}}]{Fricke:1995zz}%
\BibitemOpen
\bibfield {author} {\bibinfo {author} {\bibfnamefont {G.}~\bibnamefont
{Fricke}}, \bibinfo {author} {\bibfnamefont {C.}~\bibnamefont {Bernhardt}},
\bibinfo {author} {\bibfnamefont {K.}~\bibnamefont {Heilig}}, \bibinfo
{author} {\bibfnamefont {L.~A.}\ \bibnamefont {Schaller}}, \bibinfo {author}
{\bibfnamefont {L.}~\bibnamefont {Schellenberg}}, \bibinfo {author}
{\bibfnamefont {E.~B.}\ \bibnamefont {Shera}}, \ and\ \bibinfo {author}
{\bibfnamefont {C.~W.}\ \bibnamefont {de~Jager}},\ }\href {\doibase
10.1006/adnd.1995.1007} {\bibfield {journal} {\bibinfo {journal} {Atom.
Data Nucl. Data Tabl.}\ }\textbf {\bibinfo {volume} {60}},\ \bibinfo {pages}
{177} (\bibinfo {year} {1995})}\BibitemShut {NoStop}%
\bibitem [{\citenamefont {Bender}\ \emph {et~al.}(1999)\citenamefont {Bender},
\citenamefont {Rutz}, \citenamefont {Reinhard}, \citenamefont {Maruhn},\ and\
\citenamefont {Greiner}}]{Bender:1999yt}%
\BibitemOpen
\bibfield {author} {\bibinfo {author} {\bibfnamefont {M.}~\bibnamefont
{Bender}}, \bibinfo {author} {\bibfnamefont {K.}~\bibnamefont {Rutz}},
\bibinfo {author} {\bibfnamefont {P.~G.}\ \bibnamefont {Reinhard}}, \bibinfo
{author} {\bibfnamefont {J.~A.}\ \bibnamefont {Maruhn}}, \ and\ \bibinfo
{author} {\bibfnamefont {W.}~\bibnamefont {Greiner}},\ }\href {\doibase
10.1103/PhysRevC.60.034304} {\bibfield {journal} {\bibinfo {journal} {Phys.
Rev.}\ }\textbf {\bibinfo {volume} {C60}},\ \bibinfo {pages} {034304}
(\bibinfo {year} {1999})},\ \Eprint {http://arxiv.org/abs/nucl-th/9906030}
{nucl-th/9906030 [nucl-th]} \BibitemShut {NoStop}%
\bibitem [{\citenamefont {Abrahamyan}\ \emph {et~al.}(2012)\citenamefont
{Abrahamyan} \emph {et~al.}}]{Abrahamyan:2012gp}%
\BibitemOpen
\bibfield {author} {\bibinfo {author} {\bibfnamefont {S.}~\bibnamefont
{Abrahamyan}} \emph {et~al.} (\bibinfo {collaboration} {PREX}),\ }\href
{\doibase 10.1103/PhysRevLett.108.112502} {\bibfield {journal} {\bibinfo
{journal} {Phys. Rev. Lett.}\ }\textbf {\bibinfo {volume} {108}},\ \bibinfo
{pages} {112502} (\bibinfo {year} {2012})},\ \Eprint
{http://arxiv.org/abs/arXiv:1201.2568} {arXiv:1201.2568 [nucl-ex]}
\BibitemShut {NoStop}%
\bibitem [{\citenamefont {Horowitz}\ \emph {et~al.}(2012)\citenamefont
{Horowitz} \emph {et~al.}}]{Horowitz:2012tj}%
\BibitemOpen
\bibfield {author} {\bibinfo {author} {\bibfnamefont {C.~J.}\ \bibnamefont
{Horowitz}} \emph {et~al.},\ }\href {\doibase 10.1103/PhysRevC.85.032501}
{\bibfield {journal} {\bibinfo {journal} {Phys. Rev.}\ }\textbf {\bibinfo
{volume} {C85}},\ \bibinfo {pages} {032501} (\bibinfo {year} {2012})},\
\Eprint {http://arxiv.org/abs/arXiv:1202.1468} {arXiv:1202.1468 [nucl-ex]}
\BibitemShut {NoStop}%
\bibitem [{\citenamefont {Baker}\ and\ \citenamefont
{Cousins}(1984)}]{Baker:1983tu}%
\BibitemOpen
\bibfield {author} {\bibinfo {author} {\bibfnamefont {S.}~\bibnamefont
{Baker}}\ and\ \bibinfo {author} {\bibfnamefont {R.~D.}\ \bibnamefont
{Cousins}},\ }\href {\doibase 10.1016/0167-5087(84)90016-4} {\bibfield
{journal} {\bibinfo {journal} {Nucl. Instrum. Meth.}\ }\textbf {\bibinfo
{volume} {221}},\ \bibinfo {pages} {437} (\bibinfo {year}
{1984})}\BibitemShut {NoStop}%
\bibitem [{\citenamefont {Novales-Sanchez}\ \emph {et~al.}(2008)\citenamefont
{Novales-Sanchez}, \citenamefont {Rosado}, \citenamefont {Santiago-Olan},\
and\ \citenamefont {Toscano}}]{NovalesSanchez:2008tn}%
\BibitemOpen
\bibfield {author} {\bibinfo {author} {\bibfnamefont {H.}~\bibnamefont
{Novales-Sanchez}}, \bibinfo {author} {\bibfnamefont {A.}~\bibnamefont
{Rosado}}, \bibinfo {author} {\bibfnamefont {V.}~\bibnamefont
{Santiago-Olan}}, \ and\ \bibinfo {author} {\bibfnamefont {J.}~\bibnamefont
{Toscano}},\ }\href {\doibase 10.1103/PhysRevD.78.073014} {\bibfield
{journal} {\bibinfo {journal} {Phys. Rev.}\ }\textbf {\bibinfo {volume}
{D78}},\ \bibinfo {pages} {073014} (\bibinfo {year} {2008})},\ \Eprint
{http://arxiv.org/abs/arXiv:0805.4177} {arXiv:0805.4177 [hep-ph]}
\BibitemShut {NoStop}%
\bibitem [{\citenamefont {Akimov}\ \emph
{et~al.}(2018{\natexlab{b}})\citenamefont {Akimov} \emph
{et~al.}}]{Akimov:2018ghi}%
\BibitemOpen
\bibfield {author} {\bibinfo {author} {\bibfnamefont {D.}~\bibnamefont
{Akimov}} \emph {et~al.} (\bibinfo {collaboration} {COHERENT}),\ }\href@noop
{} {\ (\bibinfo {year} {2018}{\natexlab{b}})},\ \Eprint
{http://arxiv.org/abs/arXiv:1803.09183} {arXiv:1803.09183 [physics.ins-det]}
\BibitemShut {NoStop}%
\bibitem [{\citenamefont {Maneschg}(2018)}]{Maneschg-Neutrino2018}%
\BibitemOpen
\bibfield {author} {\bibinfo {author} {\bibfnamefont {W.}~\bibnamefont
{Maneschg}} (\bibinfo {collaboration} {CONUS}),\ }\href@noop {} {\ (\bibinfo
{year} {2018})},\ \bibinfo {note} {talk presented at {Neutrino 2018, XXVIII
International Conference on Neutrino Physics and Astrophysics, 4-9 June 2018,
Heidelberg, Germany, http://doi.org/10.5281/zenodo.1286927}}\BibitemShut
{NoStop}%
\bibitem [{\citenamefont {Aguilar-Arevalo}\ \emph {et~al.}(2016)\citenamefont
{Aguilar-Arevalo} \emph {et~al.}}]{Aguilar-Arevalo:2016khx}%
\BibitemOpen
\bibfield {author} {\bibinfo {author} {\bibfnamefont {A.}~\bibnamefont
{Aguilar-Arevalo}} \emph {et~al.} (\bibinfo {collaboration} {CONNIE}),\
}\href@noop {} {\bibfield {journal} {\bibinfo {journal} {J. Phys. Conf.
Ser.}\ }\textbf {\bibinfo {volume} {761}},\ \bibinfo {pages} {012057}
(\bibinfo {year} {2016})},\ \Eprint {http://arxiv.org/abs/arXiv:1608.01565}
{arXiv:1608.01565 [physics]} \BibitemShut {NoStop}%
\bibitem [{\citenamefont {Agnolet}\ \emph {et~al.}(2017)\citenamefont {Agnolet}
\emph {et~al.}}]{Agnolet:2016zir}%
\BibitemOpen
\bibfield {author} {\bibinfo {author} {\bibfnamefont {G.}~\bibnamefont
{Agnolet}} \emph {et~al.} (\bibinfo {collaboration} {MINER}),\ }\href@noop {}
{\bibfield {journal} {\bibinfo {journal} {Nucl.Instrum.Meth.}\ }\textbf
{\bibinfo {volume} {A853}},\ \bibinfo {pages} {53} (\bibinfo {year}
{2017})},\ \Eprint {http://arxiv.org/abs/arXiv:1609.02066} {arXiv:1609.02066
[physics]} \BibitemShut {NoStop}%
\bibitem [{\citenamefont {Strauss}\ \emph {et~al.}(2017)\citenamefont {Strauss}
\emph {et~al.}}]{Strauss:2017cuu}%
\BibitemOpen
\bibfield {author} {\bibinfo {author} {\bibfnamefont {R.}~\bibnamefont
{Strauss}} \emph {et~al.},\ }\href@noop {} {\bibfield {journal} {\bibinfo
{journal} {Eur.Phys.J.}\ }\textbf {\bibinfo {volume} {C77}},\ \bibinfo
{pages} {506} (\bibinfo {year} {2017})},\ \Eprint
{http://arxiv.org/abs/arXiv:1704.04320} {arXiv:1704.04320 [physics.ins-det]}
\BibitemShut {NoStop}%
\bibitem [{\citenamefont {Singh}\ and\ \citenamefont
{Wong}(2017)}]{Singh:2017jow}%
\BibitemOpen
\bibfield {author} {\bibinfo {author} {\bibfnamefont {L.}~\bibnamefont
{Singh}}\ and\ \bibinfo {author} {\bibfnamefont {H.~T.}\ \bibnamefont {Wong}}
(\bibinfo {collaboration} {TEXONO}),\ }\href {\doibase
10.1088/1742-6596/888/1/012124} {\bibfield {journal} {\bibinfo {journal}
{J. Phys. Conf. Ser.}\ }\textbf {\bibinfo {volume} {888}},\ \bibinfo {pages}
{012124} (\bibinfo {year} {2017})}\BibitemShut {NoStop}%
\end{thebibliography}

\end{document}